\documentclass[aps,prd,reprint,balancelastpage,nofootinbib,preprintnumbers,amsmath,amssymb,superscriptaddress]{revtex4-1}
\usepackage[
colorlinks=true,        
citecolor=blue,         
linkcolor=blue,         
urlcolor=blue           
]{hyperref}             
\usepackage{orcidlink}
\usepackage[utf8]{inputenc} 
\usepackage{graphicx}
\usepackage{bm}         

\newcommand{\nc}{\newcommand*} 
\nc{\Eq}[1]{Eq.~\eqref{#1}}     
\nc{\Fig}[1]{Fig.~\ref{#1}}     
\nc{\Table}[1]{Table~\ref{#1}}  
\nc{\Sec}[1]{Sec.~\ref{#1}}     
\nc{\red}[1]{\textcolor{red}{#1}}
\nc{\dt}{\delta}
\nc{\bt}{\mathbf{t}}
\nc{\eg}{\textit{e.g.~}}
\nc{\bn}[1]{\dt\bm{t}_{\text{#1}}}
\nc{\be}{\bm{\epsilon}}
\nc{\BF}{\mathcal{BF}}
\nc{\yr}{\mathrm{yr}}
\newcommand{\papertitle}{Constraining inflation with nonminimal derivative coupling with the Parkes Pulsar Timing Array third data release}

\def\e{\begin{equation}}
\def\q{\end{equation}}

\begin{document}
	
\title{\papertitle}

\author{Chang Han\orcidlink{0009-0009-6246-1885}}
\affiliation{Department of Physics and Synergetic Innovation Center for Quantum Effects and Applications, Hunan Normal University, Changsha, Hunan 410081, China}
\affiliation{Institute of Interdisciplinary Studies, Hunan Normal University, Changsha, Hunan 410081, China}

\author{Li-Yang Chen\orcidlink{0009-0008-5120-7016}}
\affiliation{College of Physics and Engineering Technology, Chengdu Normal University, Chengdu, Sichuan 611130, China}
\affiliation{Department of Physics and Synergetic Innovation Center for Quantum Effects and Applications, Hunan Normal University, Changsha, Hunan 410081, China}
\affiliation{Institute of Interdisciplinary Studies, Hunan Normal University, Changsha, Hunan 410081, China}

\author{Zu-Cheng Chen\orcidlink{0000-0001-7016-9934}}
\email{zuchengchen@hunnu.edu.cn}
\affiliation{Department of Physics and Synergetic Innovation Center for Quantum Effects and Applications, Hunan Normal University, Changsha, Hunan 410081, China}
\affiliation{Institute of Interdisciplinary Studies, Hunan Normal University, Changsha, Hunan 410081, China}

\author{Chengjie Fu\orcidlink{0000-0003-1651-5677}}
\email{fucj@ahnu.edu.cn}
\affiliation{Department of Physics, Anhui Normal University, Wuhu, Anhui 241002, China}

\author{Puxun Wu\orcidlink{0000-0002-9188-7393}}
\email{pxwu@hunnu.edu.cn}
\affiliation{Department of Physics and Synergetic Innovation Center for Quantum Effects and Applications, Hunan Normal University, Changsha, Hunan 410081, China}
\affiliation{Institute of Interdisciplinary Studies, Hunan Normal University, Changsha, Hunan 410081, China}

\author{Hongwei Yu\orcidlink{0000-0002-3303-9724}}
\email{hwyu@hunnu.edu.cn}
\affiliation{Department of Physics and Synergetic Innovation Center for Quantum Effects and Applications, Hunan Normal University, Changsha, Hunan 410081, China}
\affiliation{Institute of Interdisciplinary Studies, Hunan Normal University, Changsha, Hunan 410081, China}

\author{N. D. Ramesh Bhat\orcidlink{0000-0002-8383-5059}}
\affiliation{International Centre for Radio Astronomy Research, Curtin University, Bentley, WA 6102, Australia}

\author{Xiaojin Liu\orcidlink{0000-0002-2187-4087}}
\affiliation{Department of Physics, Faculty of Arts and Sciences, Beijing Normal University, Zhuhai 519087, China}

\author{Valentina Di Marco\orcidlink{0000-0003-3432-0494}}
\affiliation{School of Physics and Astronomy, Monash University, Clayton VIC 3800, Australia}
\affiliation{OzGrav: The ARC Center of Excellence for Gravitational Wave Discovery, Clayton VIC 3800, Australia}
\affiliation{Australia Telescope National Facility, CSIRO, Space and Astronomy, PO Box 76, Epping, NSW 1710, Australia}

\author{Saurav Mishra\orcidlink{0009-0001-5633-3512}}
\affiliation{Centre for Astrophysics and Supercomputing, Swinburne University of Technology, P.O. Box 218, Hawthorn, VIC 3122, Australia}

\author{Daniel J. Reardon\orcidlink{0000-0002-2035-4688}}
\affiliation{Centre for Astrophysics and Supercomputing, Swinburne University of Technology, P.O. Box 218, Hawthorn, VIC 3122, Australia}
\affiliation{OzGrav: The ARC Centre of Excellence for Gravitational Wave Discovery, Hawthorn VIC 3122, Australia}

\author{Christopher J. Russell\orcidlink{0000-0002-1942-7296}}
\affiliation{CSIRO Scientific Computing, Australian Technology Park, Locked Bag 9013, Alexandria, NSW 1435, Australia}

\author{Ryan M. Shannon\orcidlink{0000-0002-7285-6348}}
\affiliation{Centre for Astrophysics and Supercomputing, Swinburne University of Technology, P.O. Box 218, Hawthorn, VIC 3122, Australia}
\affiliation{OzGrav: The ARC Centre of Excellence for Gravitational Wave Discovery, Hawthorn VIC 3122, Australia}

\author{Lei Zhang\orcidlink{0000-0001-8539-4237}}
\affiliation{National Astronomical Observatories, Chinese Academy of Sciences, A20 Datun Road, Chaoyang District, Beijing 100101, China}
\affiliation{Centre for Astrophysics and Supercomputing, Swinburne University of Technology, P.O. Box 218, Hawthorn, VIC 3122, Australia}

\author{Xingjiang Zhu\orcidlink{0000-0001-7049-6468}}
\affiliation{Department of Physics, Faculty of Arts and Sciences, Beijing Normal University, Zhuhai 519087, China}
\affiliation{Institute for Frontier in Astronomy and Astrophysics, Beijing Normal University, Beijing 102206, China}

\author{Andrew Zic\orcidlink{0000-0002-9583-2947}}
\affiliation{Australia Telescope National Facility, CSIRO, Space and Astronomy, PO Box 76, Epping, NSW 1710, Australia}
\affiliation{OzGrav: The ARC Centre of Excellence for Gravitational Wave Discovery, Hawthorn VIC 3122, Australia}

\collaboration{The PPTA Collaboration}

\begin{abstract}
We study an inflation model with nonminimal derivative coupling that features a coupling between the  derivative  of the inflaton field and the Einstein tensor.  
This model naturally amplifies curvature perturbations at small scales via gravitationally enhanced friction, a mechanism critical for the formation of primordial black holes and the associated production of potentially detectable scalar-induced gravitational waves.
We derive analytical expressions for the primordial power spectrum, enabling efficient exploration of the model parameter space without requiring computationally intensive numerical solutions of the Mukhanov-Sasaki equation. Using the third data release of the Parkes Pulsar Timing Array (PPTA DR3), we constrain the model parameters characterizing the coupling function: $\phi_c = 3.7^{+0.3}_{-0.5} M_\mathrm{P}$, $\log_{10} \omega_L = 7.1^{+0.6}_{-0.3}$, and $\log_{10} \sigma = -8.3^{+0.3}_{-0.6}$ at 90\% confidence level. Our results demonstrate the growing capability of pulsar timing arrays to probe early Universe physics, complementing traditional cosmic microwave background observations by providing unique constraints on inflationary dynamics at small scales.

\end{abstract}
\maketitle
\section{Introduction}

The inflationary paradigm has become a cornerstone of modern cosmology, providing elegant solutions to several fundamental puzzles, including the horizon, flatness, and monopole problems, in the early Universe~\cite{Guth:1980zm,Linde:1981mu,Albrecht:1982wi,Guth:1982ec}. While the standard slow-roll (SR) inflation successfully explains the observed cosmic microwave background (CMB) anisotropies~\cite{Lyth:1998xn,Planck:2018jri}, there is growing interest in inflation models that can produce enhanced curvature perturbations at small scales. Such enhancement is particularly crucial for the formation of primordial black holes (PBHs)~\cite{Hawking:1971ei,Carr:1974nx,Garcia-Bellido:2017mdw, Sasaki:2018dmp,Braglia:2020eai}, which have emerged as compelling candidates for dark matter and potential sources of gravitational wave (GW) events detected by LIGO-Virgo-KAGRA Collaboration~\cite{Bird:2016dcv,Sasaki:2016jop,Chen:2024dxh}. The possibility of simultaneously explaining dark matter through PBHs and generating observable GW signatures provides a powerful probe of early Universe physics~\cite{Seto:2004zu,Saito:2008jc}.

Pulsar Timing Arrays (PTAs) have opened a new frontier in testing early Universe scenarios. In 2023, major PTA collaborations, including the European PTA (EPTA) in combination with Indian PTA (InPTA)~\cite{EPTA:2023sfo,Antoniadis:2023ott}, the North American Nanohertz Observatory for GWs (NANOGrav)~\cite{NANOGrav:2023hde,NANOGrav:2023gor}, the Parkes PTA (PPTA)~\cite{Zic:2023gta,Reardon:2023gzh}, and the Chinese PTA (CPTA)~\cite{Xu:2023wog}, independently reported evidence for a stochastic GW background (SGWB) with Hellings-Downs spatial correlations~\cite{Hellings:1983fr}.
Most recently, the MeerKAT PTA (MPTA) has also reported similar results~\cite{Miles:2024rjc,Miles:2024seg}.
These observations reveal a common-spectrum process with a characteristic strain amplitude of approximately $\mathcal{O}(10^{-15})$ at a reference frequency of $1\,\mathrm{yr}^{-1}$. While this signal is commonly attributed to GWs from supermassive black hole binaries (SMBHBs)~\cite{NANOGrav:2023hfp,Ellis:2023dgf,Bi:2023tib}, it may alternatively arise from primordial sources~\cite{NANOGrav:2023hvm,EPTA:2023xxk,Wu:2023hsa,Ellis:2023oxs,Figueroa:2023zhu}, offering a complementary window into inflationary dynamics.

Large-amplitude curvature perturbations at small scales not only result in PBHs formation but also generate significant scalar-induced GWs (SIGWs)~\cite{Ananda:2006af,Baumann:2007zm,Kohri:2018awv}, which have been extensively studied as potential explanations for the PTA signal~\cite{Chen:2019xse,Franciolini:2023pbf,Liu:2023ymk,Wang:2023ost,Yi:2023mbm,Tagliazucchi:2023dai,Liu:2023pau,Balaji:2023ehk,Liu:2023hpw,Wang:2023sij,Zhu:2023gmx,You:2023rmn,HosseiniMansoori:2023mqh,Yi:2023tdk,Yi:2023npi,Harigaya:2023pmw,Chen:2024fir,Chen:2024twp}. Various inflationary models for amplifying small-scale curvature perturbations have been proposed, based on mechanisms such as ultra-slow-roll (USR) inflation~\cite{Kinney:1997ne,Inoue:2001zt,Kinney:2005vj}, reduced sound speed~\cite{Kamenshchik:2018sig,Ballesteros:2018wlw}, and parametric resonance~\cite{Cai:2018tuh,Chen:2019zza,Zhou:2020kkf}. Among these, a novel model~\cite{Fu:2019ttf,Fu:2019vqc} achieves USR inflation through the mechanism of gravitationally enhanced friction, which originates from a field derivative coupling with the Einstein tensor (known as nonminimal derivative coupling)~\cite{Amendola:1993uh,Kaloper:2003yf,Sushkov:2009hk,Germani:2010gm,Germani:2011ua,Tsujikawa:2012mk,Gialamas:2020vto,Gialamas:2024jeb}. A notable feature of this model is that the enhanced curvature perturbations exhibit an approximately broken power-law spectrum, with spectral indices that can be expressed analytically in terms of the model parameters~\cite{Fu:2019vqc}. This analytical tractability is particularly valuable, as linking specific models to observational data is often hindered by computational challenges that limit systematic parameter space exploration.

In this work, we use the PPTA third data release (DR3) to constrain the inflationary model with nonminimal derivative coupling proposed in~\cite{Fu:2019ttf}. We derive an analytical expression for the primordial power spectrum to facilitate efficient parameter estimation. 
The paper is organized as follows. In \Sec{NDC}, we develop the theoretical framework, presenting both analytical expressions for the primordial power spectrum and detailed calculations of the resulting SIGW energy density spectrum. \Sec{data} describes our methodology for analyzing the PPTA DR3 data. \Sec{conclusion} presents our results and discusses their implications for early Universe physics. Technical details are provided in \textit{Supplementary Material}.

\section{Inflation with nonminimal derivative coupling}\label{NDC}

In this section, we investigate the model of inflation with nonminimal derivative coupling capable of generating enhanced primordial curvature perturbations, which subsequently lead to observable SIGWs. We begin by establishing the theoretical framework and analyzing the background evolution equations. We then derive analytical expressions for the primordial power spectrum and calculate the resulting GW energy density spectrum.

The action for the inflation model with nonminimal derivative coupling is given by~\cite{Fu:2019ttf}
\e\label{eq:action}
S=\int d^4x\sqrt{-g}\bigg[ \frac{1}{2\kappa ^2}R-\frac{1}{2}(g^{\mu \nu}-\kappa^2 \theta G^{\mu \nu }) \partial_\mu\phi \partial_\nu \phi -V(\phi) \bigg] ,
\q
where $\kappa^{-1}\equiv M_\mathrm{P} =2.4\times 10^{18}\,\mathrm{GeV} $ is the reduced Planck mass. The $\phi$ is the scalar field value and the coupling function $\theta$ is chosen to provide localized enhanced friction between the inflaton field and gravity, and takes the form~\cite{Fu:2019ttf}
\begin{equation}\label{eq:thetaphi}
\theta=\frac{\omega }{\sqrt{\kappa ^2\left ( \frac{\phi-\phi_c}{\sigma }  \right )^2+1 } }.
\end{equation}
Here, $\omega$ is a dimensionless parameter determining the maximum amplitude of the coupling, $\sigma$ is a dimensionless parameter controlling the width of the coupling region, and $\phi_c$ is a mass scale parameter specifying the field value where the coupling reaches its maximum strength.
This functional form is particularly chosen because it creates a smooth, localized enhancement of friction near $\phi=\phi_c$ that allows the inflaton to naturally transition from SR to USR and back. Such a controlled transition is essential for generating the sharp peak in the power spectrum needed for PBH formation while maintaining consistency with CMB observations at large scales.

In the spatially flat Friedmann-Robertson-Walker (FRW) spacetime, the background evolution is governed by
\begin{equation}\label{eq:HV}
       3H^2= \kappa^2\left[\frac{1}{2}\left(1+9\kappa^2 H^2\theta \right)\dot{\phi}^2+V\right],
\end{equation}
\begin{align}\label{eq:dotphi}
\left(1+3\kappa^2 H^2\theta\right)&\ddot{\phi}+3H\left[1+\kappa^2(3H^2+2\dot{H})\theta\right]\dot{\phi} \nonumber \\
&+\frac{3}{2}\kappa^2\theta_{,\phi}H^2\dot\phi^2+V_{,\phi}= 0,
\end{align}
where $H \equiv \dot{a}/a$ is the Hubble parameter, and $a = a_\mathrm{end}e^N$ is the scale factor, with $a_\mathrm{end}$ being its value at the end of inflation. Note that we define the e-folding number $N$ to count backward from the end of inflation ($N = 0$), with negative values corresponding to earlier times during the inflationary phase. This choice of convention  simplifies the tracking of perturbation evolution from horizon exit to the end of inflation.
The specific form of coupling function $\theta$ given in \Eq{eq:thetaphi} is crucial in generating a region of enhanced gravitational friction near $\phi=\phi_c$. Within this region, the inflaton field experiences substantially stronger friction compared to conventional SR regions, naturally transitioning into an USR stage of inflation. This mechanism produces a significant enhancement in the curvature perturbation power spectrum during the USR stage. As illustrated in \Fig{fig:Pkcase2}, the resulting power spectrum exhibits three distinct stages: an initial SR stage with standard amplitude, a transition period, and an USR stage characterized by substantial enhancement.

\subsection{Primordial power spectrum}\label{spectrum}
We employ a fractional power-law potential 
\begin{equation}
\label{eq:Vphi}
    V=\lambda M_\mathrm{P}^{4-p}|\phi|^p
\end{equation}
with $p=2/5$~\cite{Silverstein:2008sg}, which ensures compatibility with current CMB constraints on the scalar spectral index at large scales~\cite{Fu:2019vqc,Fu:2019ttf}. Under the SR and USR approximation, the curvature perturbation power spectrum takes the analytical form~\cite{Fu:2019ttf}
\begin{align}
\label{eq:spectrum}
    \mathcal{P_R}(\phi)\simeq &\frac{\lambda}{12\pi^2p^2}\left ( \frac{\phi}{M_\mathrm{P}}  \right )^{2+p} \nonumber \\
&\times \left[ 1+\frac{\omega \lambda }{\sqrt{\kappa ^2\left ( \frac{\phi-\phi_c}{\sigma }  \right )^2+1 } }
\left( \frac{\phi}{M_\mathrm{P}}  \right )^p   \right].
\end{align}
This power spectrum can be approximated by a broken power-law form:
\begin{align}
\mathcal{P_R}(k)\simeq\left\{\begin{array}{lcl}
k^{n_1},\;(k<k_p)\\
k^{n_2},\;(k>k_p)\end{array}\right.
\end{align}
with the spectral indices given by
\begin{equation}\label{eq:ns}
\begin{aligned}
n_1&=3\left ( 1-\sqrt{1-\frac{4}{15}(\kappa \phi_c)^{-7/5}(\omega \lambda \sigma )^-1 }  \right ),\\
n_2&=3\left ( 1-\sqrt{1+\frac{4}{15}(\kappa \phi_c)^{-7/5}(\omega \lambda \sigma )^-1 }  \right ).
\end{aligned}
\end{equation}
The broken power-law approximation matches well with the results obtained by numerically solving Mukhanov-Sasaki equation~\cite{Fu:2019vqc}.

The calculation of the SIGW energy spectrum using \Eq{eq:OmegaGW} in the following requires an explicit expression for the curvature power spectrum $\mathcal{P}_\mathcal{R}$ as a function of wavenumber $k$. This necessitates establishing a mapping between the inflaton field value $\phi$ and $k$. Traditionally, this mapping has been obtained by numerically solving the Mukhanov-Sasaki equation~\cite{Mukhanov:2005sc}:
\begin{equation}
    u_k''+(c_s^2k^2-\frac{z''}{z})u_k=0,
\end{equation}
where $c_s$ is sound speed and $u_k \equiv z\mathcal{R}_k$ (see \textit{Supplementary Material} for a detailed derivation). 
While this numerical approach provides accurate results, it is computationally intensive, typically requiring several minutes of computation time for each set of model parameters. This computational overhead makes systematic exploration of the parameter space impractical. To address this limitation, we develop an analytical approximation for the $\phi$-$k$ mapping below.

First, let us discuss the relationship between the inflaton field $\phi$ and the e-folding number $N$. By combining the results obtained in~\cite{Fu:2019vqc}, one can derive an approximate solution for Eqs.~\eqref{eq:HV} and \eqref{eq:dotphi} (see \textit{Supplementary Material} for details), expressed as:
\begin{equation}
\label{eq:H/dotphi}
    \frac{H}{\dot{\phi } }\simeq 
\left\{\begin{matrix}
-\frac{\kappa ^2V(1+\kappa^4V\theta)}{V_{,\phi}},
&\hspace{-1cm}\text{for SR stage},
\\
-\frac{\kappa ^2 V \left ( 3(1+\kappa ^4V\theta)+ \sqrt{6\theta _{,\phi}V_{,\phi}+9(1+\kappa ^4V\theta)^2}  \right ) }{6V_{,\phi}},
\\
&\hspace{-2cm} {\text{for Transition stage}},
\\
-\frac{\kappa ^6V^2\theta}{V_{,\phi}}, 
& \hspace{-1cm}{\text{for USR stage}}.
\end{matrix}\right. 
\end{equation}
Since the e-folding number $N$ follows the relation,
\begin{equation}
\label{eq:N}
      N=\int^{t}_{t_{\mathrm{end}}} H(\tilde{t}) d\tilde{t}=\int^{\phi}_{\phi_{\mathrm{end}}} \frac{H}{\dot \phi}(\tilde{\phi})d\tilde{\phi},
\end{equation}
where $\phi_{\mathrm{end}}$ can be neglected for large-field inflation, a mapping from $\phi$ to $N$ can be derived. After performing the necessary approximations and integration, one can obtain the function of $N(\phi)$ (see \textit{Supplementary Material} for details). \Fig{fig:Ncase2} demonstrates the excellent agreement between our analytical approximation and numerical results.

\begin{figure}[tbp]
\centering
\includegraphics[width=0.48\textwidth]{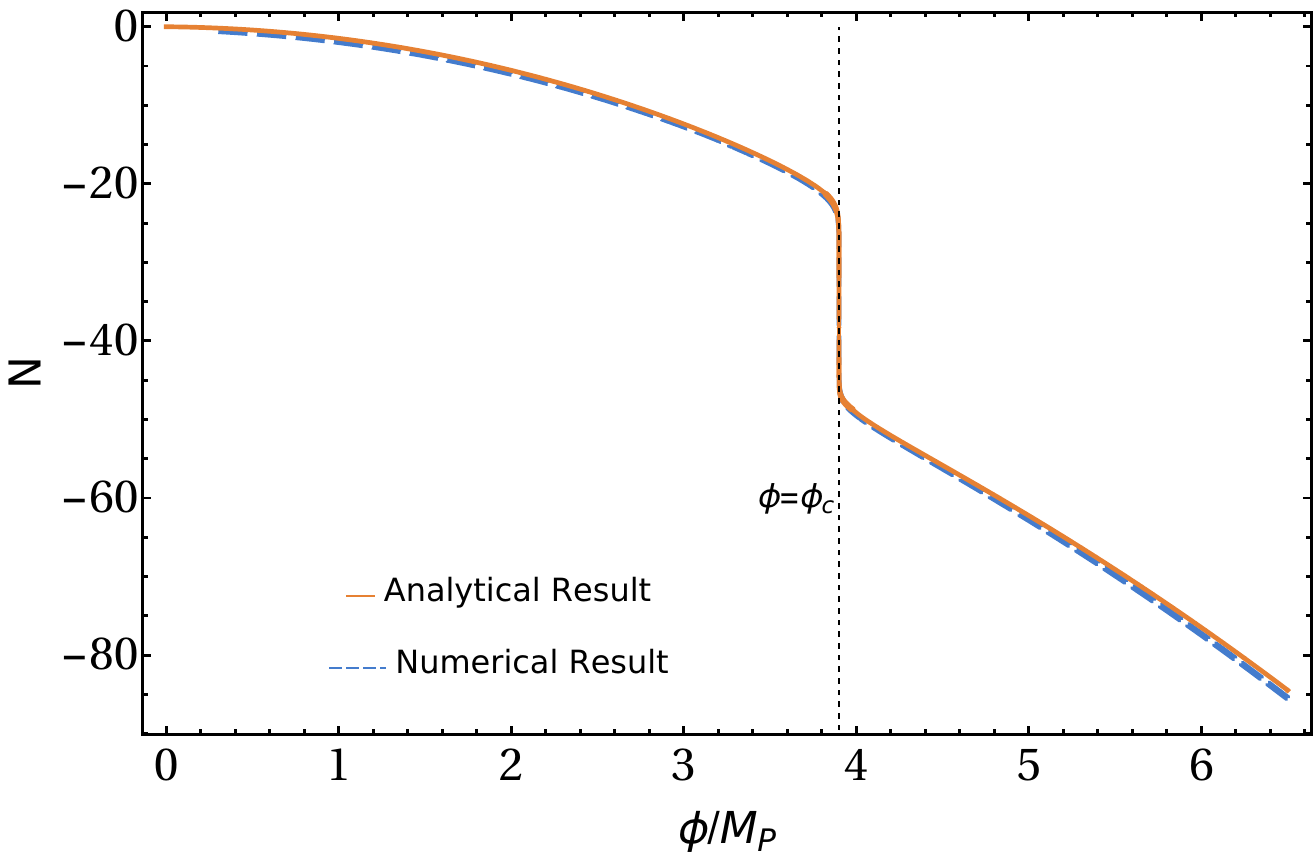}
\caption{\label{fig:Ncase2}Relationship between the e-folding number $N$ and the inflaton field value $\phi$ for parameters $\phi_c/M_\mathrm{P}=3.9$, $\omega\lambda=1.53\times10^7$, and $\sigma=3\times10^{-9}$. The orange line shows our analytical solution $N(\phi)$, while the blue line represents the numerical result from solving the Mukhanov-Sasaki equations. The vertical dashed line indicates $\phi=\phi_c$.}
\end{figure}

\begin{figure}[tbp]
\centering
\includegraphics[width=0.48\textwidth]{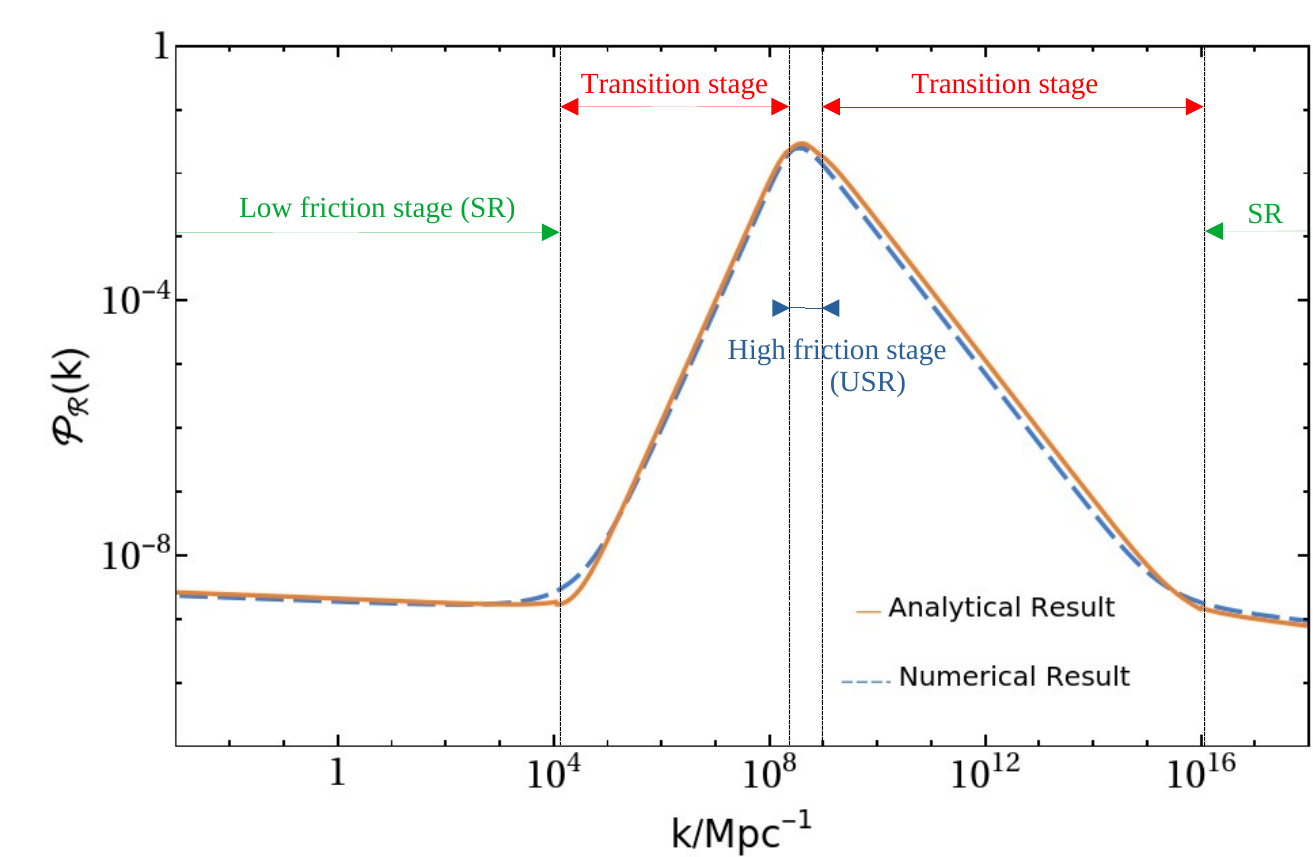}
\caption{\label{fig:Pkcase2}Primordial power spectrum $\mathcal{P_R}(k)$ for the same parameter values as in \Fig{fig:Ncase2}. The orange line shows our analytical result, while the blue line represents the numerical solution of the Mukhanov-Sasaki equations.}
\end{figure}

Using the horizon-crossing condition $c_s k=a H$ (where $c_s \simeq 1$), we can establish the relationship between the comoving wavenumber $k$ and the field value $\phi$ through
\begin{equation}
\label{eq:kHa}
k(\phi)=H(\phi)\, a(\phi)
=H(\phi)\, {a_\mathrm{end}} \, \exp\!\left[\int^{\phi}_{\phi_{\mathrm{end}}} \frac{H}{\dot \phi}(\tilde{\phi})\, d\tilde{\phi}\right].
\end{equation}
The complete expressions and detailed derivation are provided in \textit{Supplementary Material}.
By inverting this relation to obtain $\phi(k)$ and substituting into \Eq{eq:spectrum}, we obtain the power spectrum as a function of $k$. \Fig{fig:Pkcase2} demonstrates that this analytical result (orange line) closely matches the numerical solution of the Mukhanov-Sasaki equation (blue line) for representative parameters $\phi_c=3.9\, M_\mathrm{P}$, $\omega\lambda=1.53\times10^7$, and $\sigma=3\times10^{-9}$, which proves that our expression is valid and can replace the numerical solution process. 
Furthermore, the derived spectral indices are consistent with those reported in~\cite{Fu:2019vqc}.

\subsection{Scalar-induced gravitational wave}\label{SIGW}
SIGWs are generated by second-order effects when primordial curvature perturbations reenter the horizon during the radiation-dominated era. We consider perturbations to a FRW metric in the conformal Newtonian gauge~\cite{Ananda:2006af}:
\begin{equation}
       ds^2=a^2\left \{ -(1+2\Psi)d\tau^2+\left [(1-2\Psi)\delta _{ij} +\frac{1}{2}h_{ij}
       \right ] dx^idx^j \right \},
\end{equation}
where $\tau$ is conformal time, $\Psi$ denotes the Bardeen potential characterizing scalar perturbations, and $h_{ij}$ represents tensor perturbations.

After reheating, the tensor perturbations are governed by the Einstein equations, with the spatial components yielding
\begin{equation}
       h''_{ij}+2\mathcal{H}h'_{ij}-\nabla ^2h_{ij}=-4T^{lm}_{ij}S_{lm},
\end{equation}
where primes denote derivatives with respect to $\tau$, $\mathcal{H}\equiv a'/a$ is the conformal Hubble parameter, and $T^{lm}_{ij}$ is the transverse-traceless projection operator~\cite{Ananda:2006af}. The source term $S_{ij}$ is given by
\begin{equation}
       S_{ij} = 4\Psi \partial_i \partial_j \Psi + 2 \partial_i \Psi \partial_j \Psi - \frac{1}{\mathcal{H}^2} \partial_i (\mathcal{H}\Psi + \Psi') \partial_j (\mathcal{H}\Psi + \Psi').
\end{equation}

For a given comoving wave number $k$, the scalar perturbation $\Psi(\tau)$ satisfies
\begin{equation}
       \Psi''_k +\frac{4}{\tau} \Psi'_k+\frac{k^2}{3} \Psi_k=0.
\end{equation}
An analytical solution is
\begin{equation}
       \Psi_k(\tau)=\psi_k\frac{9}{(k\tau)^2}\left ( \frac{\sin\frac{k\tau}{\sqrt{3} } }{\frac{k\tau}{\sqrt{3} }} -\cos\frac{k\tau}{\sqrt{3} }\right ),
\end{equation}
where $\psi_k$ is the primordial perturbation. Therefore the power spectrum of curvature perturbations could be expressed as
\begin{equation}
      \left \langle \psi_k\psi_{\tilde{k} } \right \rangle  =\frac{2\pi^2}{k^3}\frac{4}{9}
      \mathcal{P_R}(k) \ \delta (k+\tilde{k})  .
\end{equation}
The energy density of the SIGW could be related with power spectrum as~\cite{Kohri:2018awv}
\begin{equation}
\label{eq:OmegaGW}
\begin{aligned}
    \Omega_{\mathrm{GW}}&(\tau_c, k) = \frac{1}{12} \int_0^{\infty} d\nu \int_{\lvert 1-\nu \rvert}^{1+\nu} du \left( \frac{4\nu^2 - (1 + \nu^2 - u^2)^2}{4\nu u} \right)^2 \\ 
    & \quad \times \mathcal{P_R}(uk) \mathcal{P_R}(\nu k)\left( \frac{3}{4u^3 \nu^3} \right)^2(-3 + u^2 + \nu^2)^2 \\
    &\quad \times \Bigg\{ \left[ -4\nu u + (-3 + u^2 + \nu^2) \ln \left\lvert \frac{3 - (u + \nu)^2}{3 - (u - \nu)^2} \right\rvert \right]^2  \\
    &\qquad\quad + \pi^2 (-3 + u^2 + \nu^2)^2 \Theta (v + u - \sqrt{3}) \Bigg\},
\end{aligned}
\end{equation}
where $\Theta$ is the Heaviside theta function, and $\tau_c$ is the time of the SIGW stop grow. 
The integral in Eq.~\eqref{eq:OmegaGW} is evaluated numerically using the analytical form of the power spectrum $\mathcal{P_R}(k)$ obtained in the previous subsection.
As the GW propagates to the present time, the energy density decreases as ~\cite{Inomata:2018epa}
\begin{equation}
\label{eq:OGW}
       \Omega_{\mathrm{GW}}h^2=0.83\left(\frac{g_c}{10.75}\right)^{-\frac{1}{3}}\Omega_{r,0}h^2\Omega_{\mathrm{GW}}(\tau_c,k),
\end{equation}
where $g_c\simeq 106.75$ represents the effective relativistic degrees of freedom at $\tau_c$, and $\Omega_{r,0}h^2\simeq 4.2\times10^{-5}$ is the present radiation density parameter. The frequency-wavenumber relation is given by
\begin{equation}
       f = \frac{k}{2\pi a_0} = 1.546\times 10^{-15} \left(\frac{k}{\mathrm{Mpc}^{-1}}\right) \mathrm{Hz},  
\end{equation}
where $a_0$ is the present scale factor (set to 1). This relation accounts for the redshift of the GWs from their generation during the radiation-dominated era to the present time. The numerical factor incorporates both the conversion from comoving wavenumber to physical frequency and the cosmic expansion history.

\begin{figure}[tbp]
\centering
\includegraphics[width=0.48\textwidth]{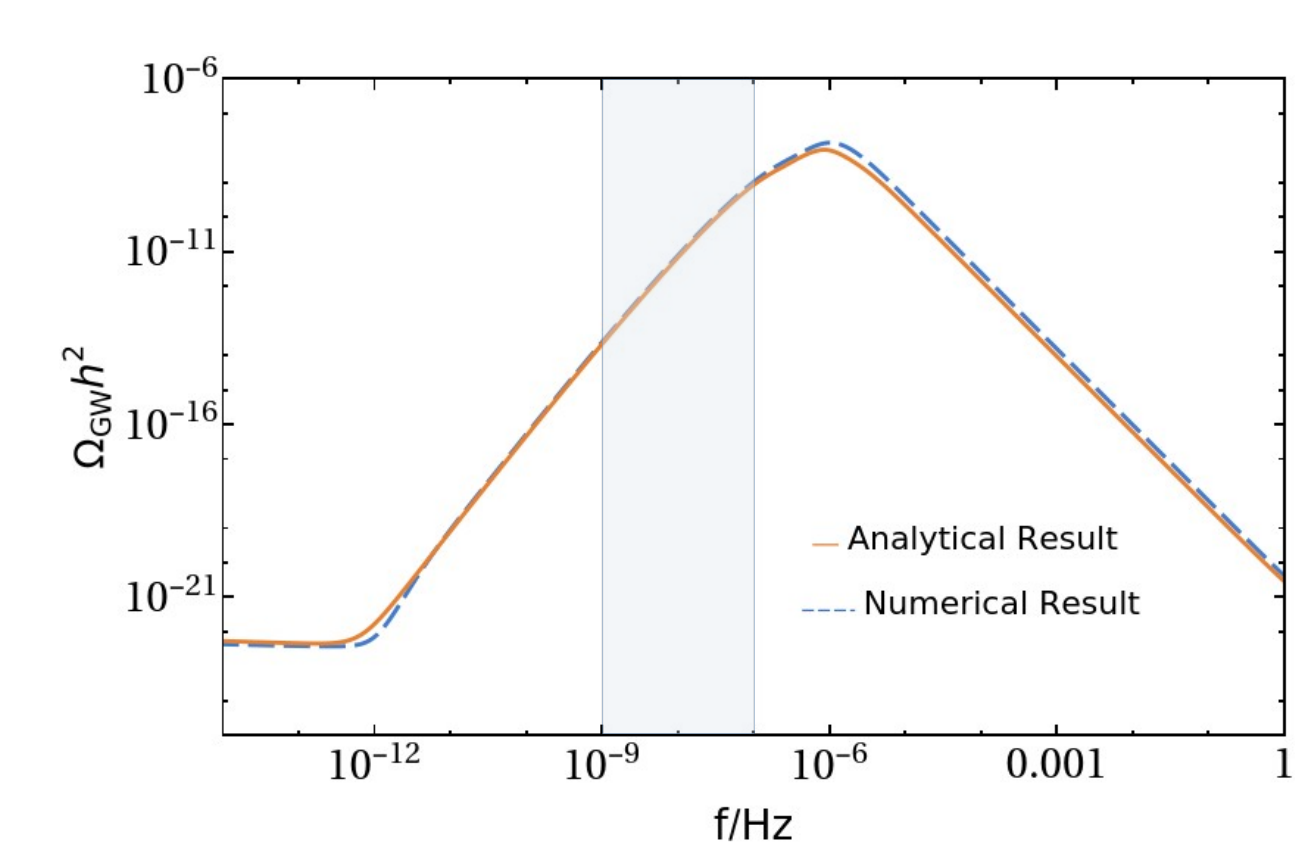}
\caption{\label{fig:OGWcase2}Present-day energy density spectrum of SIGWs, $\Omega_{\mathrm{GW}}h^2$, as a function of frequency. Results are shown for model parameters $\phi_c=3.9\, M_\mathrm{P}$, $\omega\lambda=1.53\times10^7$, and $\sigma=3\times10^{-9}$. The orange line represents calculations using our analytical expression for $\mathcal{P_R}(k)$, while the blue line shows the full numerical solution. The light blue shaded region indicates the frequency range accessible to PTA observations ($\sim 10^{-9}$--$10^{-7}$ Hz).}
\end{figure}

The energy density of SIGWs, $\Omega_{\mathrm{GW}}$, exhibits a quadratic dependence on the primordial power spectrum $\mathcal{P_R}$. Consequently, the enhancement of scalar perturbations through nonminimal derivative coupling leads to a significant amplification of the SIGW signal. 
A distinctive feature of the GWB spectrum is its characteristic peak, which emerges from modes that exit the horizon during the USR phase when $\phi \approx \phi_c$. The location of this peak can be determined analytically, with the peak wavenumber given by
\begin{equation}
    k_p \simeq \left(\frac{\lambda\phi_c^p}{3}\right)^{1/2} a_\mathrm{end}e^{N_1+N_2+\frac{1}{2}N_3},
\end{equation}
where $N_1$, $N_2$, and $N_3$ characterize the duration of different evolutionary phases (explicit expressions are provided in the \textit{Supplementary Material}). For the model parameters used in \Fig{fig:OGWcase2}, we obtain $k_p \simeq 4.7\times10^8\,\mathrm{Mpc}^{-1}$, corresponding to a peak frequency of $f_p \simeq 7.3\times10^{-7}\,\mathrm{Hz}$ in the present-day universe.
As shown in \Fig{fig:OGWcase2}, we compare the SIGW energy density spectrum calculated using \Eq{eq:OGW} for both our analytical solution of $\mathcal{P_R}(k)$ (orange line) and the full numerical computation (blue line). The excellent agreement between these results within the PTA-sensitive frequency range ($\sim 10^{-9}$--$10^{-7}$ Hz) validates our analytical approach for predicting observable SIGW signals. In the PTA-sensitive frequency range, the relative error of $\Omega_{\mathrm{GW}}h^2$ between analytical and numerical results remains at $\lesssim \mathcal{O}(10\%)$, and notably decreases with increasing $\phi_c$. This analytical formulation represents a significant improvement over previous work~\cite{Liu:2020oqe}, where relative errors could exceed an order of magnitude. Moreover, the relative error in the power-law index $n_s \equiv d\ln \Omega_\mathrm{GW}/d\ln f$ remains at $\lesssim \mathcal{O}(1\%)$ across all model parameter choices in the PTA-sensitive range.

Our nonminimal derivative coupling model is characterized by three parameters: $\phi_c$, $\omega_L\equiv \omega\lambda$, and $\sigma$. The physical consistency of the model imposes the constraint $\frac{4}{15}(\kappa\phi_c/M_\mathrm{P})^{-1-p}(\sigma\omega_L)^{-1}<1$, which avoids imaginary solutions and significantly restricts the available parameter space. Additional theoretical and observational constraints require $\frac{12}{25}(\kappa\phi_c/M_\mathrm{P})^{-1-p}(\sigma\omega_L)^{-1}>1$ and $3<\phi_c/M_\mathrm{P}<4$, ensuring consistency with CMB observations while maintaining SIGW amplitudes within PTA-detectable ranges. The normalization parameter $\lambda$ is conventionally determined by the CMB-scale power spectrum at $k^*\simeq0.05\,\mathrm{Mpc}^{-1}$, where observations indicate $\mathcal{P_R}(k^*)\simeq2.10\times10^{-9}$. In the standard SR inflation without coupling, $\lambda\approx3.76\times10^{-10}$. However, the presence of nonminimal derivative coupling induces a slight enhancement of the spectrum at CMB scales and affects the evolution of the scalar field, necessitating a modification of $\lambda$. We adopt $\lambda=8.0\times10^{-10}$ to accommodate this enhancement while maintaining consistency with observations.

\section{Data analysis}\label{data}

We use the PPTA DR3~\cite{Zic:2023gta} to constrain parameters of our nonminimal derivative coupling inflation model. PPTA DR3 encompasses $18$\,yr of high-precision timing observations from $2004$ to $2022$. The dataset includes $32$ millisecond pulsars, of which $30$ are used in this analysis after excluding PSR~J1824$-$2452A (due to strong intrinsic red noise) and PSR~J1741$+$1351 (insufficient observations) following~\cite{Reardon:2023gzh}. The observations were conducted using the Parkes 64-meter radio telescope (now known as `Murriyang'), with recent significant enhancement through the ultrawide-band low (UWL) receiver system. This new instrumentation provides improved timing precision through wider bandwidth observations ($704\sim 4032$\,MHz). The data release extends approximately $3$\,yr beyond PPTA DR2~\cite{Kerr:2020qdo}, maintaining typical observing cadences of 3 weeks~\cite{Zic:2023gta}. The timing precision varies across the pulsar sample, with the best-timed pulsars achieving root-mean-square residuals of $\sim$100\,ns.

The search for GW signals in PTA data requires a comprehensive description of stochastic processes affecting pulse arrival times. Following detailed noise analyses of individual PPTA pulsars~\cite{Goncharov:2020krd,Reardon:2023zen}, we model multiple noise components including both time-correlated (red) and uncorrelated (white) processes. The white noise model incorporates a scale parameter on the time-of-arrival uncertainties (EFAC), an added variance (EQUAD), and a per-epoch variance (ECORR) for each backend/receiver system~\cite{NANOGrav:2015aud}. Time-correlated noise is modeled using power-law processes for spin noise and dispersion measure variations. Additional components, such as, chromatic effects (scattering variations, band-specific noise), system-specific instrumental effects, and environmental contributions (solar wind variations, interstellar medium effects, magnetospheric events) are included~\cite{Reardon:2023zen}. These noise components are implemented using Bayesian inference with Gaussian processes through the \texttt{Enterprise}~\cite{enterprise} software package.

The SGWB manifests as a common process across all pulsars, characterized by the cross-power spectral density~\cite{Thrane:2013oya}:
\begin{equation}
S_{I J}(f)=\frac{H_{0}^{2}}{16 \pi^{4} f^{5}} \Gamma_{I J}(\xi)\, \Omega_{\mathrm{GW}}(f),
\end{equation}
where $H_0 = 67.4\, \mathrm{km}\,\mathrm{s}^{-1}\,\mathrm{Mpc}^{-1}$ is the Hubble constant~\cite{Planck:2018vyg}. Here, $\Gamma_{IJ}$ represents the Hellings-Downs correlation coefficients~\cite{Hellings:1983fr} between pulsars $I$ and $J$ given by
\begin{equation}
\Gamma_{I J}=\frac{3}{2}\left(\frac{1-\cos \xi}{2}\right) \ln \frac{1-\cos \xi}{2}-\frac{1-\cos \xi}{8}+\frac{1}{2},
\end{equation}
where $\xi$ is the angular separation between pulsars $I$ and $J$.
We model the common process using $15$ frequency components. In the nonminimal derivative coupling model, $\Omega_{\mathrm{GW}}(f)$ is characterized by three parameters: coupling position $\phi_c$, coupling peak height $\omega_L$, and coupling smoothing scale $\sigma$. Based on theoretical constraints discussed in \Sec{SIGW}, we adopt uniform priors: $\phi_c/M_\mathrm{P} \in [3, 4]$, $\log_{10} \omega_L \in [4, 8]$, and $\log_{10} \sigma \in [-10, -6]$.

For model selection, we compute Bayes factors defined as:
\begin{equation}
\BF \equiv \frac{\rm{Pr}(\mathcal{D}|\mathcal{M}_2)}{\rm{Pr}(\mathcal{D}|\mathcal{M}_1)},
\end{equation}
where $\rm{Pr}(\mathcal{D}|\mathcal{M})$ represents the evidence for model $\mathcal{M}$ given data $\mathcal{D}$. Following standard interpretation~\cite{BF}, $\BF \le 3$ indicates the evidence for model $\mathcal{M}_2$ over $\mathcal{M}_1$ is ``not worth more than a bare mention". We implement all Bayesian analyses using \texttt{Enterprise}\cite{enterprise} and \texttt{Enterprise\_extensions}~\cite{enterprise_ext} through the \texttt{PTArcade}~\cite{Mitridate:2023oar,andrea_mitridate_2023} wrapper, with Bayes factors estimated via the product-space method~\cite{10.2307/2346151,10.2307/1391010,Hee:2015eba,Taylor:2020zpk}.

\section{Results and Discussion}\label{conclusion}

\begin{figure}[tbp]
\centering
\includegraphics[width=0.48\textwidth]{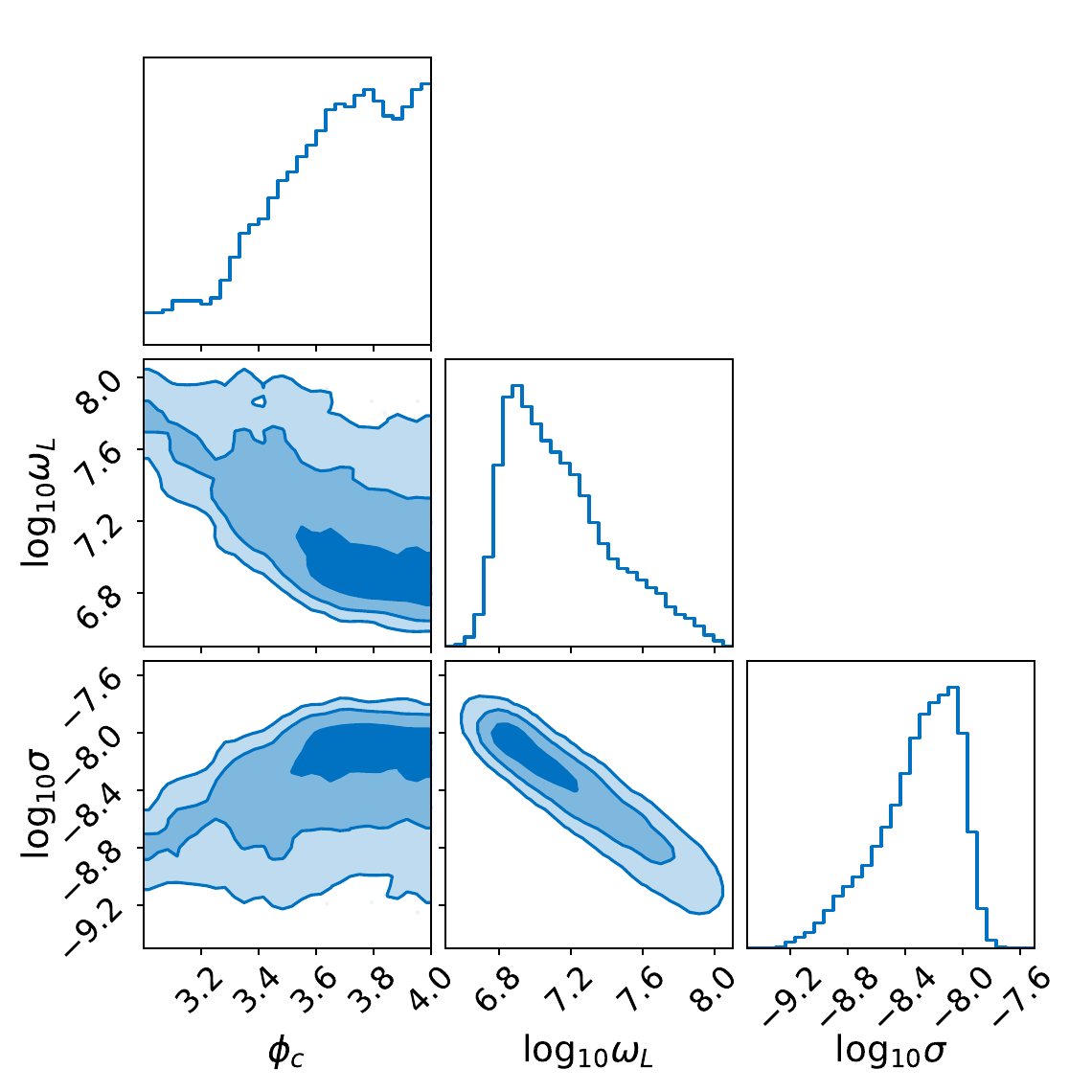}
\caption{\label{posts_ppta}Posterior distributions for the model parameters $\{\phi_c, \omega_L, \sigma\}$ constrained by PPTA DR3. The contours represent the $1\sigma$, $2\sigma$, and $3\sigma$ confidence levels in the two-dimensional plots.}
\end{figure}

\begin{figure}[t!]
\centering
\includegraphics[width=0.48\textwidth]{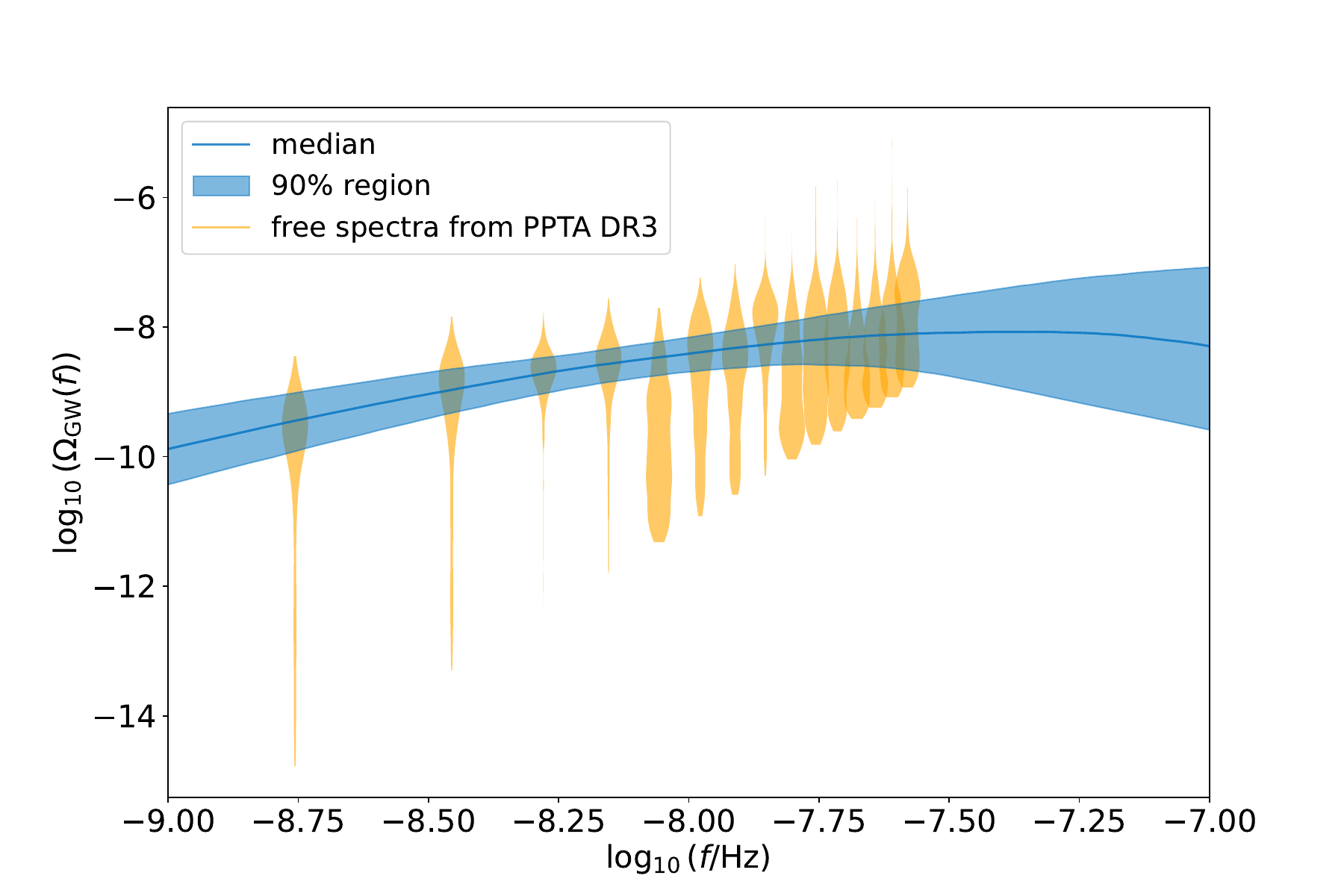}
\caption{\label{ogw_PPD}Comparison between the predicted SIGW energy density spectrum from our inflationary model and PPTA DR3 observations. The solid blue line shows the median prediction from the posterior distribution, with the shaded region indicating the $90\%$ credible interval. Orange violin plots represent the free spectra obtained from the PPTA DR3~\cite{Reardon:2023gzh}.}
\end{figure}

Using PPTA DR3, we constrain the parameters of our nonminimal derivative coupling inflation model. \Fig{posts_ppta} shows the posterior distributions for the model parameters, yielding $\phi_c = 3.7^{+0.3}_{-0.5}\,M_\mathrm{P}$, $\log_{10} \omega_L = 7.1^{+0.6}_{-0.3}$, and $\log_{10} \sigma = -8.3^{+0.3}_{-0.6}$ (quoted as median values with $90\%$ credible intervals). These constraints demonstrate that PTA observations can effectively probe the shape of the coupling function $\theta(\phi)$, with $\phi_c$ determining its position, $\omega_L$ its peak height, and $\sigma$ its smoothing scale. Note that theoretical considerations and CMB observations require $3 < \phi_c/M_\mathrm{P} < 4$ to ensure viable inflation while maintaining SIGW amplitudes within PTA-detectable ranges. Our analysis of the PPTA DR3 data provides a lower limit $\phi_c \gtrsim 3.2\,M_\mathrm{P}$ at 95\% credible level.

To assess the model's viability, we compare it with the SMBHB hypothesis, where the GW energy density spectrum follows a power-law:
\begin{equation}
\Omega_{\rm{PL}}(f) =\frac{2\pi^2A_{\rm{PL}}^2}{3H_0^2}\left(\frac{f}{f_{\yr}}\right)^{5-\gamma_{\rm{PL}}}f_{\yr}^2,
\end{equation}
with $A_{\rm{PL}}$ being the characteristic strain amplitude at $f_{\yr}=1/\rm{yr}$ and $\gamma_{\rm{PL}}=13/3$ the expected spectral index. The Bayes factor between our model and the power-law model is $0.9$, indicating that current sensitivity of PPTA data is insufficient to distinguish between these scenarios.

\Fig{ogw_PPD} presents the posterior predictive distribution for the SIGW energy density spectrum from our model. The median prediction (blue line) and $90\%$ credible region (shaded area) are consistent with the free spectrum estimates from PPTA DR3 (orange violins), supporting the viability of our model.

Our analysis demonstrates a couple of key advances. First, we have developed an analytical approximation for the primordial curvature power spectrum in nonminimal derivative coupling inflation, avoiding computationally expensive numerical solutions of the Mukhanov-Sasaki equations. This approximation reduces computation time from minutes per parameter set to effectively instantaneous calculations while maintaining accuracy, enabling efficient exploration of the parameter space. Second, we have demonstrated that PTA observations can effectively constrain inflationary models with enhanced small-scale power spectra. This establishes PTAs as a powerful complement to CMB observations, extending our ability to probe inflationary dynamics across an unprecedented range of scales.

\begin{acknowledgments}
This work has been carried out by the Parkes Pulsar Timing Array, which is part of the International Pulsar Timing Array. The Parkes radio telescope (``Murriyang'') is part of the Australia Telescope, which is funded by the Commonwealth Government for operation as a National Facility managed by CSIRO. This work was supported in part by the National Key Research and Development Program of China Grant No.~2020YFC2201502;
the National Natural Science Foundation of China under Grants  No.~12405056, No.~12305057, No.~12275080, No.~12203004, No.~12103069 and No.~12075084;
the innovative research group of Hunan Province under Grant No.~2024JJ1006;
and the Fundamental Research Funds for the Central Universities.
VDM is supported via the Australian Research Council (ARC) Centre of Excellence CE170100004 and CE230100016 and receives support from the Australian Government Research Training Program. 
\end{acknowledgments}

\bibliography{ref}

\begin{thebibliography}{93}%
\makeatletter
\providecommand \@ifxundefined [1]{%
 \@ifx{#1\undefined}
}%
\providecommand \@ifnum [1]{%
 \ifnum #1\expandafter \@firstoftwo
 \else \expandafter \@secondoftwo
 \fi
}%
\providecommand \@ifx [1]{%
 \ifx #1\expandafter \@firstoftwo
 \else \expandafter \@secondoftwo
 \fi
}%
\providecommand \natexlab [1]{#1}%
\providecommand \enquote  [1]{``#1''}%
\providecommand \bibnamefont  [1]{#1}%
\providecommand \bibfnamefont [1]{#1}%
\providecommand \citenamefont [1]{#1}%
\providecommand \href@noop [0]{\@secondoftwo}%
\providecommand \href [0]{\begingroup \@sanitize@url \@href}%
\providecommand \@href[1]{\@@startlink{#1}\@@href}%
\providecommand \@@href[1]{\endgroup#1\@@endlink}%
\providecommand \@sanitize@url [0]{\catcode `\\12\catcode `\$12\catcode `\&12\catcode `\#12\catcode `\^12\catcode `\_12\catcode `\%12\relax}%
\providecommand \@@startlink[1]{}%
\providecommand \@@endlink[0]{}%
\providecommand \url  [0]{\begingroup\@sanitize@url \@url }%
\providecommand \@url [1]{\endgroup\@href {#1}{\urlprefix }}%
\providecommand \urlprefix  [0]{URL }%
\providecommand \Eprint [0]{\href }%
\providecommand \doibase [0]{http://dx.doi.org/}%
\providecommand \selectlanguage [0]{\@gobble}%
\providecommand \bibinfo  [0]{\@secondoftwo}%
\providecommand \bibfield  [0]{\@secondoftwo}%
\providecommand \translation [1]{[#1]}%
\providecommand \BibitemOpen [0]{}%
\providecommand \bibitemStop [0]{}%
\providecommand \bibitemNoStop [0]{.\EOS\space}%
\providecommand \EOS [0]{\spacefactor3000\relax}%
\providecommand \BibitemShut  [1]{\csname bibitem#1\endcsname}%
\let\auto@bib@innerbib\@empty
\bibitem [{\citenamefont {Guth}(1981)}]{Guth:1980zm}%
  \BibitemOpen
  \bibfield  {author} {\bibinfo {author} {\bibfnamefont {A.~H.}\ \bibnamefont {Guth}},\ }\href {\doibase 10.1103/PhysRevD.23.347} {\bibfield  {journal} {\bibinfo  {journal} {Phys. Rev. D}\ }\textbf {\bibinfo {volume} {23}},\ \bibinfo {pages} {347} (\bibinfo {year} {1981})}\BibitemShut {NoStop}%
\bibitem [{\citenamefont {Linde}(1982)}]{Linde:1981mu}%
  \BibitemOpen
  \bibfield  {author} {\bibinfo {author} {\bibfnamefont {A.~D.}\ \bibnamefont {Linde}},\ }\href {\doibase 10.1016/0370-2693(82)91219-9} {\bibfield  {journal} {\bibinfo  {journal} {Phys. Lett. B}\ }\textbf {\bibinfo {volume} {108}},\ \bibinfo {pages} {389} (\bibinfo {year} {1982})}\BibitemShut {NoStop}%
\bibitem [{\citenamefont {Albrecht}\ and\ \citenamefont {Steinhardt}(1982)}]{Albrecht:1982wi}%
  \BibitemOpen
  \bibfield  {author} {\bibinfo {author} {\bibfnamefont {A.}~\bibnamefont {Albrecht}}\ and\ \bibinfo {author} {\bibfnamefont {P.~J.}\ \bibnamefont {Steinhardt}},\ }\href {\doibase 10.1103/PhysRevLett.48.1220} {\bibfield  {journal} {\bibinfo  {journal} {Phys. Rev. Lett.}\ }\textbf {\bibinfo {volume} {48}},\ \bibinfo {pages} {1220} (\bibinfo {year} {1982})}\BibitemShut {NoStop}%
\bibitem [{\citenamefont {Guth}\ and\ \citenamefont {Pi}(1982)}]{Guth:1982ec}%
  \BibitemOpen
  \bibfield  {author} {\bibinfo {author} {\bibfnamefont {A.~H.}\ \bibnamefont {Guth}}\ and\ \bibinfo {author} {\bibfnamefont {S.~Y.}\ \bibnamefont {Pi}},\ }\href {\doibase 10.1103/PhysRevLett.49.1110} {\bibfield  {journal} {\bibinfo  {journal} {Phys. Rev. Lett.}\ }\textbf {\bibinfo {volume} {49}},\ \bibinfo {pages} {1110} (\bibinfo {year} {1982})}\BibitemShut {NoStop}%
\bibitem [{\citenamefont {Lyth}\ and\ \citenamefont {Riotto}(1999)}]{Lyth:1998xn}%
  \BibitemOpen
  \bibfield  {author} {\bibinfo {author} {\bibfnamefont {D.~H.}\ \bibnamefont {Lyth}}\ and\ \bibinfo {author} {\bibfnamefont {A.}~\bibnamefont {Riotto}},\ }\href {\doibase 10.1016/S0370-1573(98)00128-8} {\bibfield  {journal} {\bibinfo  {journal} {Phys. Rept.}\ }\textbf {\bibinfo {volume} {314}},\ \bibinfo {pages} {1} (\bibinfo {year} {1999})},\ \Eprint {http://arxiv.org/abs/hep-ph/9807278} {arXiv:hep-ph/9807278} \BibitemShut {NoStop}%
\bibitem [{\citenamefont {Akrami}\ \emph {et~al.}(2020)\citenamefont {Akrami} \emph {et~al.}}]{Planck:2018jri}%
  \BibitemOpen
  \bibfield  {author} {\bibinfo {author} {\bibfnamefont {Y.}~\bibnamefont {Akrami}} \emph {et~al.} (\bibinfo {collaboration} {Planck}),\ }\href {\doibase 10.1051/0004-6361/201833887} {\bibfield  {journal} {\bibinfo  {journal} {Astron. Astrophys.}\ }\textbf {\bibinfo {volume} {641}},\ \bibinfo {pages} {A10} (\bibinfo {year} {2020})},\ \Eprint {http://arxiv.org/abs/1807.06211} {arXiv:1807.06211 [astro-ph.CO]} \BibitemShut {NoStop}%
\bibitem [{\citenamefont {Hawking}(1971)}]{Hawking:1971ei}%
  \BibitemOpen
  \bibfield  {author} {\bibinfo {author} {\bibfnamefont {S.}~\bibnamefont {Hawking}},\ }\href {\doibase 10.1093/mnras/152.1.75} {\bibfield  {journal} {\bibinfo  {journal} {Mon. Not. Roy. Astron. Soc.}\ }\textbf {\bibinfo {volume} {152}},\ \bibinfo {pages} {75} (\bibinfo {year} {1971})}\BibitemShut {NoStop}%
\bibitem [{\citenamefont {Carr}\ and\ \citenamefont {Hawking}(1974)}]{Carr:1974nx}%
  \BibitemOpen
  \bibfield  {author} {\bibinfo {author} {\bibfnamefont {B.~J.}\ \bibnamefont {Carr}}\ and\ \bibinfo {author} {\bibfnamefont {S.~W.}\ \bibnamefont {Hawking}},\ }\href {\doibase 10.1093/mnras/168.2.399} {\bibfield  {journal} {\bibinfo  {journal} {Mon. Not. Roy. Astron. Soc.}\ }\textbf {\bibinfo {volume} {168}},\ \bibinfo {pages} {399} (\bibinfo {year} {1974})}\BibitemShut {NoStop}%
\bibitem [{\citenamefont {Garcia-Bellido}\ and\ \citenamefont {Ruiz~Morales}(2017)}]{Garcia-Bellido:2017mdw}%
  \BibitemOpen
  \bibfield  {author} {\bibinfo {author} {\bibfnamefont {J.}~\bibnamefont {Garcia-Bellido}}\ and\ \bibinfo {author} {\bibfnamefont {E.}~\bibnamefont {Ruiz~Morales}},\ }\href {\doibase 10.1016/j.dark.2017.09.007} {\bibfield  {journal} {\bibinfo  {journal} {Phys. Dark Univ.}\ }\textbf {\bibinfo {volume} {18}},\ \bibinfo {pages} {47} (\bibinfo {year} {2017})},\ \Eprint {http://arxiv.org/abs/1702.03901} {arXiv:1702.03901 [astro-ph.CO]} \BibitemShut {NoStop}%
\bibitem [{\citenamefont {Sasaki}\ \emph {et~al.}(2018)\citenamefont {Sasaki}, \citenamefont {Suyama}, \citenamefont {Tanaka},\ and\ \citenamefont {Yokoyama}}]{Sasaki:2018dmp}%
  \BibitemOpen
  \bibfield  {author} {\bibinfo {author} {\bibfnamefont {M.}~\bibnamefont {Sasaki}}, \bibinfo {author} {\bibfnamefont {T.}~\bibnamefont {Suyama}}, \bibinfo {author} {\bibfnamefont {T.}~\bibnamefont {Tanaka}}, \ and\ \bibinfo {author} {\bibfnamefont {S.}~\bibnamefont {Yokoyama}},\ }\href {\doibase 10.1088/1361-6382/aaa7b4} {\bibfield  {journal} {\bibinfo  {journal} {Class. Quant. Grav.}\ }\textbf {\bibinfo {volume} {35}},\ \bibinfo {pages} {063001} (\bibinfo {year} {2018})},\ \Eprint {http://arxiv.org/abs/1801.05235} {arXiv:1801.05235 [astro-ph.CO]} \BibitemShut {NoStop}%
\bibitem [{\citenamefont {Braglia}\ \emph {et~al.}(2020)\citenamefont {Braglia}, \citenamefont {Hazra}, \citenamefont {Finelli}, \citenamefont {Smoot}, \citenamefont {Sriramkumar},\ and\ \citenamefont {Starobinsky}}]{Braglia:2020eai}%
  \BibitemOpen
  \bibfield  {author} {\bibinfo {author} {\bibfnamefont {M.}~\bibnamefont {Braglia}}, \bibinfo {author} {\bibfnamefont {D.~K.}\ \bibnamefont {Hazra}}, \bibinfo {author} {\bibfnamefont {F.}~\bibnamefont {Finelli}}, \bibinfo {author} {\bibfnamefont {G.~F.}\ \bibnamefont {Smoot}}, \bibinfo {author} {\bibfnamefont {L.}~\bibnamefont {Sriramkumar}}, \ and\ \bibinfo {author} {\bibfnamefont {A.~A.}\ \bibnamefont {Starobinsky}},\ }\href {\doibase 10.1088/1475-7516/2020/08/001} {\bibfield  {journal} {\bibinfo  {journal} {JCAP}\ }\textbf {\bibinfo {volume} {08}},\ \bibinfo {pages} {001} (\bibinfo {year} {2020})},\ \Eprint {http://arxiv.org/abs/2005.02895} {arXiv:2005.02895 [astro-ph.CO]} \BibitemShut {NoStop}%
\bibitem [{\citenamefont {Bird}\ \emph {et~al.}(2016)\citenamefont {Bird}, \citenamefont {Cholis}, \citenamefont {Mu\~noz}, \citenamefont {Ali-Ha\"\i{}moud}, \citenamefont {Kamionkowski}, \citenamefont {Kovetz}, \citenamefont {Raccanelli},\ and\ \citenamefont {Riess}}]{Bird:2016dcv}%
  \BibitemOpen
  \bibfield  {author} {\bibinfo {author} {\bibfnamefont {S.}~\bibnamefont {Bird}}, \bibinfo {author} {\bibfnamefont {I.}~\bibnamefont {Cholis}}, \bibinfo {author} {\bibfnamefont {J.~B.}\ \bibnamefont {Mu\~noz}}, \bibinfo {author} {\bibfnamefont {Y.}~\bibnamefont {Ali-Ha\"\i{}moud}}, \bibinfo {author} {\bibfnamefont {M.}~\bibnamefont {Kamionkowski}}, \bibinfo {author} {\bibfnamefont {E.~D.}\ \bibnamefont {Kovetz}}, \bibinfo {author} {\bibfnamefont {A.}~\bibnamefont {Raccanelli}}, \ and\ \bibinfo {author} {\bibfnamefont {A.~G.}\ \bibnamefont {Riess}},\ }\href {\doibase 10.1103/PhysRevLett.116.201301} {\bibfield  {journal} {\bibinfo  {journal} {Phys. Rev. Lett.}\ }\textbf {\bibinfo {volume} {116}},\ \bibinfo {pages} {201301} (\bibinfo {year} {2016})},\ \Eprint {http://arxiv.org/abs/1603.00464} {arXiv:1603.00464 [astro-ph.CO]} \BibitemShut {NoStop}%
\bibitem [{\citenamefont {Sasaki}\ \emph {et~al.}(2016)\citenamefont {Sasaki}, \citenamefont {Suyama}, \citenamefont {Tanaka},\ and\ \citenamefont {Yokoyama}}]{Sasaki:2016jop}%
  \BibitemOpen
  \bibfield  {author} {\bibinfo {author} {\bibfnamefont {M.}~\bibnamefont {Sasaki}}, \bibinfo {author} {\bibfnamefont {T.}~\bibnamefont {Suyama}}, \bibinfo {author} {\bibfnamefont {T.}~\bibnamefont {Tanaka}}, \ and\ \bibinfo {author} {\bibfnamefont {S.}~\bibnamefont {Yokoyama}},\ }\href {\doibase 10.1103/PhysRevLett.117.061101} {\bibfield  {journal} {\bibinfo  {journal} {Phys. Rev. Lett.}\ }\textbf {\bibinfo {volume} {117}},\ \bibinfo {pages} {061101} (\bibinfo {year} {2016})},\ \bibinfo {note} {[Erratum: Phys.Rev.Lett. 121, 059901 (2018)]},\ \Eprint {http://arxiv.org/abs/1603.08338} {arXiv:1603.08338 [astro-ph.CO]} \BibitemShut {NoStop}%
\bibitem [{\citenamefont {Chen}\ and\ \citenamefont {Hall}(2024)}]{Chen:2024dxh}%
  \BibitemOpen
  \bibfield  {author} {\bibinfo {author} {\bibfnamefont {Z.-C.}\ \bibnamefont {Chen}}\ and\ \bibinfo {author} {\bibfnamefont {A.}~\bibnamefont {Hall}},\ }\href@noop {} {\  (\bibinfo {year} {2024})},\ \Eprint {http://arxiv.org/abs/2402.03934} {arXiv:2402.03934 [astro-ph.CO]} \BibitemShut {NoStop}%
\bibitem [{\citenamefont {Seto}\ and\ \citenamefont {Cooray}(2004)}]{Seto:2004zu}%
  \BibitemOpen
  \bibfield  {author} {\bibinfo {author} {\bibfnamefont {N.}~\bibnamefont {Seto}}\ and\ \bibinfo {author} {\bibfnamefont {A.}~\bibnamefont {Cooray}},\ }\href {\doibase 10.1103/PhysRevD.70.063512} {\bibfield  {journal} {\bibinfo  {journal} {Phys. Rev. D}\ }\textbf {\bibinfo {volume} {70}},\ \bibinfo {pages} {063512} (\bibinfo {year} {2004})},\ \Eprint {http://arxiv.org/abs/astro-ph/0405216} {arXiv:astro-ph/0405216} \BibitemShut {NoStop}%
\bibitem [{\citenamefont {Saito}\ and\ \citenamefont {Yokoyama}(2009)}]{Saito:2008jc}%
  \BibitemOpen
  \bibfield  {author} {\bibinfo {author} {\bibfnamefont {R.}~\bibnamefont {Saito}}\ and\ \bibinfo {author} {\bibfnamefont {J.}~\bibnamefont {Yokoyama}},\ }\href {\doibase 10.1103/PhysRevLett.102.161101} {\bibfield  {journal} {\bibinfo  {journal} {Phys. Rev. Lett.}\ }\textbf {\bibinfo {volume} {102}},\ \bibinfo {pages} {161101} (\bibinfo {year} {2009})},\ \bibinfo {note} {[Erratum: Phys.Rev.Lett. 107, 069901 (2011)]},\ \Eprint {http://arxiv.org/abs/0812.4339} {arXiv:0812.4339 [astro-ph]} \BibitemShut {NoStop}%
\bibitem [{\citenamefont {Antoniadis}\ \emph {et~al.}(2023{\natexlab{a}})\citenamefont {Antoniadis} \emph {et~al.}}]{EPTA:2023sfo}%
  \BibitemOpen
  \bibfield  {author} {\bibinfo {author} {\bibfnamefont {J.}~\bibnamefont {Antoniadis}} \emph {et~al.} (\bibinfo {collaboration} {EPTA}),\ }\href {\doibase 10.1051/0004-6361/202346841} {\bibfield  {journal} {\bibinfo  {journal} {Astron. Astrophys.}\ }\textbf {\bibinfo {volume} {678}},\ \bibinfo {pages} {A48} (\bibinfo {year} {2023}{\natexlab{a}})},\ \Eprint {http://arxiv.org/abs/2306.16224} {arXiv:2306.16224 [astro-ph.HE]} \BibitemShut {NoStop}%
\bibitem [{\citenamefont {Antoniadis}\ \emph {et~al.}(2023{\natexlab{b}})\citenamefont {Antoniadis} \emph {et~al.}}]{Antoniadis:2023ott}%
  \BibitemOpen
  \bibfield  {author} {\bibinfo {author} {\bibfnamefont {J.}~\bibnamefont {Antoniadis}} \emph {et~al.} (\bibinfo {collaboration} {EPTA, InPTA:}),\ }\href {\doibase 10.1051/0004-6361/202346844} {\bibfield  {journal} {\bibinfo  {journal} {Astron. Astrophys.}\ }\textbf {\bibinfo {volume} {678}},\ \bibinfo {pages} {A50} (\bibinfo {year} {2023}{\natexlab{b}})},\ \Eprint {http://arxiv.org/abs/2306.16214} {arXiv:2306.16214 [astro-ph.HE]} \BibitemShut {NoStop}%
\bibitem [{\citenamefont {Agazie}\ \emph {et~al.}(2023{\natexlab{a}})\citenamefont {Agazie} \emph {et~al.}}]{NANOGrav:2023hde}%
  \BibitemOpen
  \bibfield  {author} {\bibinfo {author} {\bibfnamefont {G.}~\bibnamefont {Agazie}} \emph {et~al.} (\bibinfo {collaboration} {NANOGrav}),\ }\href {\doibase 10.3847/2041-8213/acda9a} {\bibfield  {journal} {\bibinfo  {journal} {Astrophys. J. Lett.}\ }\textbf {\bibinfo {volume} {951}},\ \bibinfo {pages} {L9} (\bibinfo {year} {2023}{\natexlab{a}})},\ \Eprint {http://arxiv.org/abs/2306.16217} {arXiv:2306.16217 [astro-ph.HE]} \BibitemShut {NoStop}%
\bibitem [{\citenamefont {Agazie}\ \emph {et~al.}(2023{\natexlab{b}})\citenamefont {Agazie} \emph {et~al.}}]{NANOGrav:2023gor}%
  \BibitemOpen
  \bibfield  {author} {\bibinfo {author} {\bibfnamefont {G.}~\bibnamefont {Agazie}} \emph {et~al.} (\bibinfo {collaboration} {NANOGrav}),\ }\href {\doibase 10.3847/2041-8213/acdac6} {\bibfield  {journal} {\bibinfo  {journal} {Astrophys. J. Lett.}\ }\textbf {\bibinfo {volume} {951}},\ \bibinfo {pages} {L8} (\bibinfo {year} {2023}{\natexlab{b}})},\ \Eprint {http://arxiv.org/abs/2306.16213} {arXiv:2306.16213 [astro-ph.HE]} \BibitemShut {NoStop}%
\bibitem [{\citenamefont {Zic}\ \emph {et~al.}(2023)\citenamefont {Zic} \emph {et~al.}}]{Zic:2023gta}%
  \BibitemOpen
  \bibfield  {author} {\bibinfo {author} {\bibfnamefont {A.}~\bibnamefont {Zic}} \emph {et~al.},\ }\href {\doibase 10.1017/pasa.2023.36} {\bibfield  {journal} {\bibinfo  {journal} {Publ. Astron. Soc. Austral.}\ }\textbf {\bibinfo {volume} {40}},\ \bibinfo {pages} {e049} (\bibinfo {year} {2023})},\ \Eprint {http://arxiv.org/abs/2306.16230} {arXiv:2306.16230 [astro-ph.HE]} \BibitemShut {NoStop}%
\bibitem [{\citenamefont {Reardon}\ \emph {et~al.}(2023{\natexlab{a}})\citenamefont {Reardon} \emph {et~al.}}]{Reardon:2023gzh}%
  \BibitemOpen
  \bibfield  {author} {\bibinfo {author} {\bibfnamefont {D.~J.}\ \bibnamefont {Reardon}} \emph {et~al.},\ }\href {\doibase 10.3847/2041-8213/acdd02} {\bibfield  {journal} {\bibinfo  {journal} {Astrophys. J. Lett.}\ }\textbf {\bibinfo {volume} {951}},\ \bibinfo {pages} {L6} (\bibinfo {year} {2023}{\natexlab{a}})},\ \Eprint {http://arxiv.org/abs/2306.16215} {arXiv:2306.16215 [astro-ph.HE]} \BibitemShut {NoStop}%
\bibitem [{\citenamefont {Xu}\ \emph {et~al.}(2023)\citenamefont {Xu} \emph {et~al.}}]{Xu:2023wog}%
  \BibitemOpen
  \bibfield  {author} {\bibinfo {author} {\bibfnamefont {H.}~\bibnamefont {Xu}} \emph {et~al.},\ }\href {\doibase 10.1088/1674-4527/acdfa5} {\bibfield  {journal} {\bibinfo  {journal} {Res. Astron. Astrophys.}\ }\textbf {\bibinfo {volume} {23}},\ \bibinfo {pages} {075024} (\bibinfo {year} {2023})},\ \Eprint {http://arxiv.org/abs/2306.16216} {arXiv:2306.16216 [astro-ph.HE]} \BibitemShut {NoStop}%
\bibitem [{\citenamefont {Hellings}\ and\ \citenamefont {Downs}(1983)}]{Hellings:1983fr}%
  \BibitemOpen
  \bibfield  {author} {\bibinfo {author} {\bibfnamefont {R.~w.}\ \bibnamefont {Hellings}}\ and\ \bibinfo {author} {\bibfnamefont {G.~s.}\ \bibnamefont {Downs}},\ }\href {\doibase 10.1086/183954} {\bibfield  {journal} {\bibinfo  {journal} {Astrophys. J. Lett.}\ }\textbf {\bibinfo {volume} {265}},\ \bibinfo {pages} {L39} (\bibinfo {year} {1983})}\BibitemShut {NoStop}%
\bibitem [{\citenamefont {Miles}\ \emph {et~al.}(2025{\natexlab{a}})\citenamefont {Miles} \emph {et~al.}}]{Miles:2024rjc}%
  \BibitemOpen
  \bibfield  {author} {\bibinfo {author} {\bibfnamefont {M.~T.}\ \bibnamefont {Miles}} \emph {et~al.},\ }\href {\doibase 10.1093/mnras/stae2572} {\bibfield  {journal} {\bibinfo  {journal} {Mon. Not. Roy. Astron. Soc.}\ }\textbf {\bibinfo {volume} {536}},\ \bibinfo {pages} {1467} (\bibinfo {year} {2025}{\natexlab{a}})},\ \Eprint {http://arxiv.org/abs/2412.01148} {arXiv:2412.01148 [astro-ph.HE]} \BibitemShut {NoStop}%
\bibitem [{\citenamefont {Miles}\ \emph {et~al.}(2025{\natexlab{b}})\citenamefont {Miles} \emph {et~al.}}]{Miles:2024seg}%
  \BibitemOpen
  \bibfield  {author} {\bibinfo {author} {\bibfnamefont {M.~T.}\ \bibnamefont {Miles}} \emph {et~al.},\ }\href {\doibase 10.1093/mnras/stae2571} {\bibfield  {journal} {\bibinfo  {journal} {Mon. Not. Roy. Astron. Soc.}\ }\textbf {\bibinfo {volume} {536}},\ \bibinfo {pages} {1489} (\bibinfo {year} {2025}{\natexlab{b}})},\ \Eprint {http://arxiv.org/abs/2412.01153} {arXiv:2412.01153 [astro-ph.HE]} \BibitemShut {NoStop}%
\bibitem [{\citenamefont {Agazie}\ \emph {et~al.}(2023{\natexlab{c}})\citenamefont {Agazie} \emph {et~al.}}]{NANOGrav:2023hfp}%
  \BibitemOpen
  \bibfield  {author} {\bibinfo {author} {\bibfnamefont {G.}~\bibnamefont {Agazie}} \emph {et~al.} (\bibinfo {collaboration} {NANOGrav}),\ }\href {\doibase 10.3847/2041-8213/ace18b} {\bibfield  {journal} {\bibinfo  {journal} {Astrophys. J. Lett.}\ }\textbf {\bibinfo {volume} {952}},\ \bibinfo {pages} {L37} (\bibinfo {year} {2023}{\natexlab{c}})},\ \Eprint {http://arxiv.org/abs/2306.16220} {arXiv:2306.16220 [astro-ph.HE]} \BibitemShut {NoStop}%
\bibitem [{\citenamefont {Ellis}\ \emph {et~al.}(2024{\natexlab{a}})\citenamefont {Ellis}, \citenamefont {Fairbairn}, \citenamefont {H\"utsi}, \citenamefont {Raidal}, \citenamefont {Urrutia}, \citenamefont {Vaskonen},\ and\ \citenamefont {Veerm\"ae}}]{Ellis:2023dgf}%
  \BibitemOpen
  \bibfield  {author} {\bibinfo {author} {\bibfnamefont {J.}~\bibnamefont {Ellis}}, \bibinfo {author} {\bibfnamefont {M.}~\bibnamefont {Fairbairn}}, \bibinfo {author} {\bibfnamefont {G.}~\bibnamefont {H\"utsi}}, \bibinfo {author} {\bibfnamefont {J.}~\bibnamefont {Raidal}}, \bibinfo {author} {\bibfnamefont {J.}~\bibnamefont {Urrutia}}, \bibinfo {author} {\bibfnamefont {V.}~\bibnamefont {Vaskonen}}, \ and\ \bibinfo {author} {\bibfnamefont {H.}~\bibnamefont {Veerm\"ae}},\ }\href {\doibase 10.1103/PhysRevD.109.L021302} {\bibfield  {journal} {\bibinfo  {journal} {Phys. Rev. D}\ }\textbf {\bibinfo {volume} {109}},\ \bibinfo {pages} {L021302} (\bibinfo {year} {2024}{\natexlab{a}})},\ \Eprint {http://arxiv.org/abs/2306.17021} {arXiv:2306.17021 [astro-ph.CO]} \BibitemShut {NoStop}%
\bibitem [{\citenamefont {Bi}\ \emph {et~al.}(2023)\citenamefont {Bi}, \citenamefont {Wu}, \citenamefont {Chen},\ and\ \citenamefont {Huang}}]{Bi:2023tib}%
  \BibitemOpen
  \bibfield  {author} {\bibinfo {author} {\bibfnamefont {Y.-C.}\ \bibnamefont {Bi}}, \bibinfo {author} {\bibfnamefont {Y.-M.}\ \bibnamefont {Wu}}, \bibinfo {author} {\bibfnamefont {Z.-C.}\ \bibnamefont {Chen}}, \ and\ \bibinfo {author} {\bibfnamefont {Q.-G.}\ \bibnamefont {Huang}},\ }\href {\doibase 10.1007/s11433-023-2252-4} {\bibfield  {journal} {\bibinfo  {journal} {Sci. China Phys. Mech. Astron.}\ }\textbf {\bibinfo {volume} {66}},\ \bibinfo {pages} {120402} (\bibinfo {year} {2023})},\ \Eprint {http://arxiv.org/abs/2307.00722} {arXiv:2307.00722 [astro-ph.CO]} \BibitemShut {NoStop}%
\bibitem [{\citenamefont {Afzal}\ \emph {et~al.}(2023)\citenamefont {Afzal} \emph {et~al.}}]{NANOGrav:2023hvm}%
  \BibitemOpen
  \bibfield  {author} {\bibinfo {author} {\bibfnamefont {A.}~\bibnamefont {Afzal}} \emph {et~al.} (\bibinfo {collaboration} {NANOGrav}),\ }\href {\doibase 10.3847/2041-8213/acdc91} {\bibfield  {journal} {\bibinfo  {journal} {Astrophys. J. Lett.}\ }\textbf {\bibinfo {volume} {951}},\ \bibinfo {pages} {L11} (\bibinfo {year} {2023})},\ \bibinfo {note} {[Erratum: Astrophys.J.Lett. 971, L27 (2024), Erratum: Astrophys.J. 971, L27 (2024)]},\ \Eprint {http://arxiv.org/abs/2306.16219} {arXiv:2306.16219 [astro-ph.HE]} \BibitemShut {NoStop}%
\bibitem [{\citenamefont {Antoniadis}\ \emph {et~al.}(2024)\citenamefont {Antoniadis} \emph {et~al.}}]{EPTA:2023xxk}%
  \BibitemOpen
  \bibfield  {author} {\bibinfo {author} {\bibfnamefont {J.}~\bibnamefont {Antoniadis}} \emph {et~al.} (\bibinfo {collaboration} {EPTA, InPTA}),\ }\href {\doibase 10.1051/0004-6361/202347433} {\bibfield  {journal} {\bibinfo  {journal} {Astron. Astrophys.}\ }\textbf {\bibinfo {volume} {685}},\ \bibinfo {pages} {A94} (\bibinfo {year} {2024})},\ \Eprint {http://arxiv.org/abs/2306.16227} {arXiv:2306.16227 [astro-ph.CO]} \BibitemShut {NoStop}%
\bibitem [{\citenamefont {Wu}\ \emph {et~al.}(2024)\citenamefont {Wu}, \citenamefont {Chen},\ and\ \citenamefont {Huang}}]{Wu:2023hsa}%
  \BibitemOpen
  \bibfield  {author} {\bibinfo {author} {\bibfnamefont {Y.-M.}\ \bibnamefont {Wu}}, \bibinfo {author} {\bibfnamefont {Z.-C.}\ \bibnamefont {Chen}}, \ and\ \bibinfo {author} {\bibfnamefont {Q.-G.}\ \bibnamefont {Huang}},\ }\href {\doibase 10.1007/s11433-023-2298-7} {\bibfield  {journal} {\bibinfo  {journal} {Sci. China Phys. Mech. Astron.}\ }\textbf {\bibinfo {volume} {67}},\ \bibinfo {pages} {240412} (\bibinfo {year} {2024})},\ \Eprint {http://arxiv.org/abs/2307.03141} {arXiv:2307.03141 [astro-ph.CO]} \BibitemShut {NoStop}%
\bibitem [{\citenamefont {Ellis}\ \emph {et~al.}(2024{\natexlab{b}})\citenamefont {Ellis}, \citenamefont {Fairbairn}, \citenamefont {Franciolini}, \citenamefont {H\"utsi}, \citenamefont {Iovino}, \citenamefont {Lewicki}, \citenamefont {Raidal}, \citenamefont {Urrutia}, \citenamefont {Vaskonen},\ and\ \citenamefont {Veerm\"ae}}]{Ellis:2023oxs}%
  \BibitemOpen
  \bibfield  {author} {\bibinfo {author} {\bibfnamefont {J.}~\bibnamefont {Ellis}}, \bibinfo {author} {\bibfnamefont {M.}~\bibnamefont {Fairbairn}}, \bibinfo {author} {\bibfnamefont {G.}~\bibnamefont {Franciolini}}, \bibinfo {author} {\bibfnamefont {G.}~\bibnamefont {H\"utsi}}, \bibinfo {author} {\bibfnamefont {A.}~\bibnamefont {Iovino}}, \bibinfo {author} {\bibfnamefont {M.}~\bibnamefont {Lewicki}}, \bibinfo {author} {\bibfnamefont {M.}~\bibnamefont {Raidal}}, \bibinfo {author} {\bibfnamefont {J.}~\bibnamefont {Urrutia}}, \bibinfo {author} {\bibfnamefont {V.}~\bibnamefont {Vaskonen}}, \ and\ \bibinfo {author} {\bibfnamefont {H.}~\bibnamefont {Veerm\"ae}},\ }\href {\doibase 10.1103/PhysRevD.109.023522} {\bibfield  {journal} {\bibinfo  {journal} {Phys. Rev. D}\ }\textbf {\bibinfo {volume} {109}},\ \bibinfo {pages} {023522} (\bibinfo {year} {2024}{\natexlab{b}})},\ \Eprint {http://arxiv.org/abs/2308.08546} {arXiv:2308.08546 [astro-ph.CO]} \BibitemShut {NoStop}%
\bibitem [{\citenamefont {Figueroa}\ \emph {et~al.}(2024)\citenamefont {Figueroa}, \citenamefont {Pieroni}, \citenamefont {Ricciardone},\ and\ \citenamefont {Simakachorn}}]{Figueroa:2023zhu}%
  \BibitemOpen
  \bibfield  {author} {\bibinfo {author} {\bibfnamefont {D.~G.}\ \bibnamefont {Figueroa}}, \bibinfo {author} {\bibfnamefont {M.}~\bibnamefont {Pieroni}}, \bibinfo {author} {\bibfnamefont {A.}~\bibnamefont {Ricciardone}}, \ and\ \bibinfo {author} {\bibfnamefont {P.}~\bibnamefont {Simakachorn}},\ }\href {\doibase 10.1103/PhysRevLett.132.171002} {\bibfield  {journal} {\bibinfo  {journal} {Phys. Rev. Lett.}\ }\textbf {\bibinfo {volume} {132}},\ \bibinfo {pages} {171002} (\bibinfo {year} {2024})},\ \Eprint {http://arxiv.org/abs/2307.02399} {arXiv:2307.02399 [astro-ph.CO]} \BibitemShut {NoStop}%
\bibitem [{\citenamefont {Ananda}\ \emph {et~al.}(2007)\citenamefont {Ananda}, \citenamefont {Clarkson},\ and\ \citenamefont {Wands}}]{Ananda:2006af}%
  \BibitemOpen
  \bibfield  {author} {\bibinfo {author} {\bibfnamefont {K.~N.}\ \bibnamefont {Ananda}}, \bibinfo {author} {\bibfnamefont {C.}~\bibnamefont {Clarkson}}, \ and\ \bibinfo {author} {\bibfnamefont {D.}~\bibnamefont {Wands}},\ }\href {\doibase 10.1103/PhysRevD.75.123518} {\bibfield  {journal} {\bibinfo  {journal} {Phys. Rev. D}\ }\textbf {\bibinfo {volume} {75}},\ \bibinfo {pages} {123518} (\bibinfo {year} {2007})},\ \Eprint {http://arxiv.org/abs/gr-qc/0612013} {arXiv:gr-qc/0612013} \BibitemShut {NoStop}%
\bibitem [{\citenamefont {Baumann}\ \emph {et~al.}(2007)\citenamefont {Baumann}, \citenamefont {Steinhardt}, \citenamefont {Takahashi},\ and\ \citenamefont {Ichiki}}]{Baumann:2007zm}%
  \BibitemOpen
  \bibfield  {author} {\bibinfo {author} {\bibfnamefont {D.}~\bibnamefont {Baumann}}, \bibinfo {author} {\bibfnamefont {P.~J.}\ \bibnamefont {Steinhardt}}, \bibinfo {author} {\bibfnamefont {K.}~\bibnamefont {Takahashi}}, \ and\ \bibinfo {author} {\bibfnamefont {K.}~\bibnamefont {Ichiki}},\ }\href {\doibase 10.1103/PhysRevD.76.084019} {\bibfield  {journal} {\bibinfo  {journal} {Phys. Rev. D}\ }\textbf {\bibinfo {volume} {76}},\ \bibinfo {pages} {084019} (\bibinfo {year} {2007})},\ \Eprint {http://arxiv.org/abs/hep-th/0703290} {arXiv:hep-th/0703290} \BibitemShut {NoStop}%
\bibitem [{\citenamefont {Kohri}\ and\ \citenamefont {Terada}(2018)}]{Kohri:2018awv}%
  \BibitemOpen
  \bibfield  {author} {\bibinfo {author} {\bibfnamefont {K.}~\bibnamefont {Kohri}}\ and\ \bibinfo {author} {\bibfnamefont {T.}~\bibnamefont {Terada}},\ }\href {\doibase 10.1103/PhysRevD.97.123532} {\bibfield  {journal} {\bibinfo  {journal} {Phys. Rev. D}\ }\textbf {\bibinfo {volume} {97}},\ \bibinfo {pages} {123532} (\bibinfo {year} {2018})},\ \Eprint {http://arxiv.org/abs/1804.08577} {arXiv:1804.08577 [gr-qc]} \BibitemShut {NoStop}%
\bibitem [{\citenamefont {Chen}\ \emph {et~al.}(2020)\citenamefont {Chen}, \citenamefont {Yuan},\ and\ \citenamefont {Huang}}]{Chen:2019xse}%
  \BibitemOpen
  \bibfield  {author} {\bibinfo {author} {\bibfnamefont {Z.-C.}\ \bibnamefont {Chen}}, \bibinfo {author} {\bibfnamefont {C.}~\bibnamefont {Yuan}}, \ and\ \bibinfo {author} {\bibfnamefont {Q.-G.}\ \bibnamefont {Huang}},\ }\href {\doibase 10.1103/PhysRevLett.124.251101} {\bibfield  {journal} {\bibinfo  {journal} {Phys. Rev. Lett.}\ }\textbf {\bibinfo {volume} {124}},\ \bibinfo {pages} {251101} (\bibinfo {year} {2020})},\ \Eprint {http://arxiv.org/abs/1910.12239} {arXiv:1910.12239 [astro-ph.CO]} \BibitemShut {NoStop}%
\bibitem [{\citenamefont {Franciolini}\ \emph {et~al.}(2023)\citenamefont {Franciolini}, \citenamefont {Iovino}, \citenamefont {Vaskonen},\ and\ \citenamefont {Veermae}}]{Franciolini:2023pbf}%
  \BibitemOpen
  \bibfield  {author} {\bibinfo {author} {\bibfnamefont {G.}~\bibnamefont {Franciolini}}, \bibinfo {author} {\bibfnamefont {A.}~\bibnamefont {Iovino}, \bibfnamefont {Junior.}}, \bibinfo {author} {\bibfnamefont {V.}~\bibnamefont {Vaskonen}}, \ and\ \bibinfo {author} {\bibfnamefont {H.}~\bibnamefont {Veermae}},\ }\href {\doibase 10.1103/PhysRevLett.131.201401} {\bibfield  {journal} {\bibinfo  {journal} {Phys. Rev. Lett.}\ }\textbf {\bibinfo {volume} {131}},\ \bibinfo {pages} {201401} (\bibinfo {year} {2023})},\ \Eprint {http://arxiv.org/abs/2306.17149} {arXiv:2306.17149 [astro-ph.CO]} \BibitemShut {NoStop}%
\bibitem [{\citenamefont {Liu}\ \emph {et~al.}(2024{\natexlab{a}})\citenamefont {Liu}, \citenamefont {Chen},\ and\ \citenamefont {Huang}}]{Liu:2023ymk}%
  \BibitemOpen
  \bibfield  {author} {\bibinfo {author} {\bibfnamefont {L.}~\bibnamefont {Liu}}, \bibinfo {author} {\bibfnamefont {Z.-C.}\ \bibnamefont {Chen}}, \ and\ \bibinfo {author} {\bibfnamefont {Q.-G.}\ \bibnamefont {Huang}},\ }\href {\doibase 10.1103/PhysRevD.109.L061301} {\bibfield  {journal} {\bibinfo  {journal} {Phys. Rev. D}\ }\textbf {\bibinfo {volume} {109}},\ \bibinfo {pages} {L061301} (\bibinfo {year} {2024}{\natexlab{a}})},\ \Eprint {http://arxiv.org/abs/2307.01102} {arXiv:2307.01102 [astro-ph.CO]} \BibitemShut {NoStop}%
\bibitem [{\citenamefont {Wang}\ \emph {et~al.}(2024{\natexlab{a}})\citenamefont {Wang}, \citenamefont {Zhao}, \citenamefont {Li},\ and\ \citenamefont {Zhu}}]{Wang:2023ost}%
  \BibitemOpen
  \bibfield  {author} {\bibinfo {author} {\bibfnamefont {S.}~\bibnamefont {Wang}}, \bibinfo {author} {\bibfnamefont {Z.-C.}\ \bibnamefont {Zhao}}, \bibinfo {author} {\bibfnamefont {J.-P.}\ \bibnamefont {Li}}, \ and\ \bibinfo {author} {\bibfnamefont {Q.-H.}\ \bibnamefont {Zhu}},\ }\href {\doibase 10.1103/PhysRevResearch.6.L012060} {\bibfield  {journal} {\bibinfo  {journal} {Phys. Rev. Res.}\ }\textbf {\bibinfo {volume} {6}},\ \bibinfo {pages} {L012060} (\bibinfo {year} {2024}{\natexlab{a}})},\ \Eprint {http://arxiv.org/abs/2307.00572} {arXiv:2307.00572 [astro-ph.CO]} \BibitemShut {NoStop}%
\bibitem [{\citenamefont {Yi}\ \emph {et~al.}(2023)\citenamefont {Yi}, \citenamefont {Gao}, \citenamefont {Gong}, \citenamefont {Wang},\ and\ \citenamefont {Zhang}}]{Yi:2023mbm}%
  \BibitemOpen
  \bibfield  {author} {\bibinfo {author} {\bibfnamefont {Z.}~\bibnamefont {Yi}}, \bibinfo {author} {\bibfnamefont {Q.}~\bibnamefont {Gao}}, \bibinfo {author} {\bibfnamefont {Y.}~\bibnamefont {Gong}}, \bibinfo {author} {\bibfnamefont {Y.}~\bibnamefont {Wang}}, \ and\ \bibinfo {author} {\bibfnamefont {F.}~\bibnamefont {Zhang}},\ }\href {\doibase 10.1007/s11433-023-2266-1} {\bibfield  {journal} {\bibinfo  {journal} {Sci. China Phys. Mech. Astron.}\ }\textbf {\bibinfo {volume} {66}},\ \bibinfo {pages} {120404} (\bibinfo {year} {2023})},\ \Eprint {http://arxiv.org/abs/2307.02467} {arXiv:2307.02467 [gr-qc]} \BibitemShut {NoStop}%
\bibitem [{\citenamefont {Tagliazucchi}\ \emph {et~al.}(2025)\citenamefont {Tagliazucchi}, \citenamefont {Braglia}, \citenamefont {Finelli},\ and\ \citenamefont {Pieroni}}]{Tagliazucchi:2023dai}%
  \BibitemOpen
  \bibfield  {author} {\bibinfo {author} {\bibfnamefont {M.}~\bibnamefont {Tagliazucchi}}, \bibinfo {author} {\bibfnamefont {M.}~\bibnamefont {Braglia}}, \bibinfo {author} {\bibfnamefont {F.}~\bibnamefont {Finelli}}, \ and\ \bibinfo {author} {\bibfnamefont {M.}~\bibnamefont {Pieroni}},\ }\href {\doibase 10.1103/PhysRevD.111.L021305} {\bibfield  {journal} {\bibinfo  {journal} {Phys. Rev. D}\ }\textbf {\bibinfo {volume} {111}},\ \bibinfo {pages} {L021305} (\bibinfo {year} {2025})},\ \Eprint {http://arxiv.org/abs/2310.08527} {arXiv:2310.08527 [astro-ph.CO]} \BibitemShut {NoStop}%
\bibitem [{\citenamefont {Liu}\ \emph {et~al.}(2023)\citenamefont {Liu}, \citenamefont {Chen},\ and\ \citenamefont {Huang}}]{Liu:2023pau}%
  \BibitemOpen
  \bibfield  {author} {\bibinfo {author} {\bibfnamefont {L.}~\bibnamefont {Liu}}, \bibinfo {author} {\bibfnamefont {Z.-C.}\ \bibnamefont {Chen}}, \ and\ \bibinfo {author} {\bibfnamefont {Q.-G.}\ \bibnamefont {Huang}},\ }\href {\doibase 10.1088/1475-7516/2023/11/071} {\bibfield  {journal} {\bibinfo  {journal} {JCAP}\ }\textbf {\bibinfo {volume} {11}},\ \bibinfo {pages} {071} (\bibinfo {year} {2023})},\ \Eprint {http://arxiv.org/abs/2307.14911} {arXiv:2307.14911 [astro-ph.CO]} \BibitemShut {NoStop}%
\bibitem [{\citenamefont {Balaji}\ \emph {et~al.}(2023)\citenamefont {Balaji}, \citenamefont {Dom\`enech},\ and\ \citenamefont {Franciolini}}]{Balaji:2023ehk}%
  \BibitemOpen
  \bibfield  {author} {\bibinfo {author} {\bibfnamefont {S.}~\bibnamefont {Balaji}}, \bibinfo {author} {\bibfnamefont {G.}~\bibnamefont {Dom\`enech}}, \ and\ \bibinfo {author} {\bibfnamefont {G.}~\bibnamefont {Franciolini}},\ }\href {\doibase 10.1088/1475-7516/2023/10/041} {\bibfield  {journal} {\bibinfo  {journal} {JCAP}\ }\textbf {\bibinfo {volume} {10}},\ \bibinfo {pages} {041} (\bibinfo {year} {2023})},\ \Eprint {http://arxiv.org/abs/2307.08552} {arXiv:2307.08552 [gr-qc]} \BibitemShut {NoStop}%
\bibitem [{\citenamefont {Liu}\ \emph {et~al.}(2024{\natexlab{b}})\citenamefont {Liu}, \citenamefont {Wu},\ and\ \citenamefont {Chen}}]{Liu:2023hpw}%
  \BibitemOpen
  \bibfield  {author} {\bibinfo {author} {\bibfnamefont {L.}~\bibnamefont {Liu}}, \bibinfo {author} {\bibfnamefont {Y.}~\bibnamefont {Wu}}, \ and\ \bibinfo {author} {\bibfnamefont {Z.-C.}\ \bibnamefont {Chen}},\ }\href {\doibase 10.1088/1475-7516/2024/04/011} {\bibfield  {journal} {\bibinfo  {journal} {JCAP}\ }\textbf {\bibinfo {volume} {04}},\ \bibinfo {pages} {011} (\bibinfo {year} {2024}{\natexlab{b}})},\ \Eprint {http://arxiv.org/abs/2310.16500} {arXiv:2310.16500 [astro-ph.CO]} \BibitemShut {NoStop}%
\bibitem [{\citenamefont {Wang}\ \emph {et~al.}(2024{\natexlab{b}})\citenamefont {Wang}, \citenamefont {Zhao},\ and\ \citenamefont {Zhu}}]{Wang:2023sij}%
  \BibitemOpen
  \bibfield  {author} {\bibinfo {author} {\bibfnamefont {S.}~\bibnamefont {Wang}}, \bibinfo {author} {\bibfnamefont {Z.-C.}\ \bibnamefont {Zhao}}, \ and\ \bibinfo {author} {\bibfnamefont {Q.-H.}\ \bibnamefont {Zhu}},\ }\href {\doibase 10.1103/PhysRevResearch.6.013207} {\bibfield  {journal} {\bibinfo  {journal} {Phys. Rev. Res.}\ }\textbf {\bibinfo {volume} {6}},\ \bibinfo {pages} {013207} (\bibinfo {year} {2024}{\natexlab{b}})},\ \Eprint {http://arxiv.org/abs/2307.03095} {arXiv:2307.03095 [astro-ph.CO]} \BibitemShut {NoStop}%
\bibitem [{\citenamefont {Zhu}\ \emph {et~al.}(2024)\citenamefont {Zhu}, \citenamefont {Zhao}, \citenamefont {Wang},\ and\ \citenamefont {Zhang}}]{Zhu:2023gmx}%
  \BibitemOpen
  \bibfield  {author} {\bibinfo {author} {\bibfnamefont {Q.-H.}\ \bibnamefont {Zhu}}, \bibinfo {author} {\bibfnamefont {Z.-C.}\ \bibnamefont {Zhao}}, \bibinfo {author} {\bibfnamefont {S.}~\bibnamefont {Wang}}, \ and\ \bibinfo {author} {\bibfnamefont {X.}~\bibnamefont {Zhang}},\ }\href {\doibase 10.1088/1674-1137/ad79d5} {\bibfield  {journal} {\bibinfo  {journal} {Chin. Phys. C}\ }\textbf {\bibinfo {volume} {48}},\ \bibinfo {pages} {125105} (\bibinfo {year} {2024})},\ \Eprint {http://arxiv.org/abs/2307.13574} {arXiv:2307.13574 [astro-ph.CO]} \BibitemShut {NoStop}%
\bibitem [{\citenamefont {You}\ \emph {et~al.}(2023)\citenamefont {You}, \citenamefont {Yi},\ and\ \citenamefont {Wu}}]{You:2023rmn}%
  \BibitemOpen
  \bibfield  {author} {\bibinfo {author} {\bibfnamefont {Z.-Q.}\ \bibnamefont {You}}, \bibinfo {author} {\bibfnamefont {Z.}~\bibnamefont {Yi}}, \ and\ \bibinfo {author} {\bibfnamefont {Y.}~\bibnamefont {Wu}},\ }\href {\doibase 10.1088/1475-7516/2023/11/065} {\bibfield  {journal} {\bibinfo  {journal} {JCAP}\ }\textbf {\bibinfo {volume} {11}},\ \bibinfo {pages} {065} (\bibinfo {year} {2023})},\ \Eprint {http://arxiv.org/abs/2307.04419} {arXiv:2307.04419 [gr-qc]} \BibitemShut {NoStop}%
\bibitem [{\citenamefont {Hosseini~Mansoori}\ \emph {et~al.}(2023)\citenamefont {Hosseini~Mansoori}, \citenamefont {Felegray}, \citenamefont {Talebian},\ and\ \citenamefont {Sami}}]{HosseiniMansoori:2023mqh}%
  \BibitemOpen
  \bibfield  {author} {\bibinfo {author} {\bibfnamefont {S.~A.}\ \bibnamefont {Hosseini~Mansoori}}, \bibinfo {author} {\bibfnamefont {F.}~\bibnamefont {Felegray}}, \bibinfo {author} {\bibfnamefont {A.}~\bibnamefont {Talebian}}, \ and\ \bibinfo {author} {\bibfnamefont {M.}~\bibnamefont {Sami}},\ }\href {\doibase 10.1088/1475-7516/2023/08/067} {\bibfield  {journal} {\bibinfo  {journal} {JCAP}\ }\textbf {\bibinfo {volume} {08}},\ \bibinfo {pages} {067} (\bibinfo {year} {2023})},\ \Eprint {http://arxiv.org/abs/2307.06757} {arXiv:2307.06757 [astro-ph.CO]} \BibitemShut {NoStop}%
\bibitem [{\citenamefont {Yi}\ \emph {et~al.}(2024{\natexlab{a}})\citenamefont {Yi}, \citenamefont {You},\ and\ \citenamefont {Wu}}]{Yi:2023tdk}%
  \BibitemOpen
  \bibfield  {author} {\bibinfo {author} {\bibfnamefont {Z.}~\bibnamefont {Yi}}, \bibinfo {author} {\bibfnamefont {Z.-Q.}\ \bibnamefont {You}}, \ and\ \bibinfo {author} {\bibfnamefont {Y.}~\bibnamefont {Wu}},\ }\href {\doibase 10.1088/1475-7516/2024/01/066} {\bibfield  {journal} {\bibinfo  {journal} {JCAP}\ }\textbf {\bibinfo {volume} {01}},\ \bibinfo {pages} {066} (\bibinfo {year} {2024}{\natexlab{a}})},\ \Eprint {http://arxiv.org/abs/2308.05632} {arXiv:2308.05632 [astro-ph.CO]} \BibitemShut {NoStop}%
\bibitem [{\citenamefont {Yi}\ \emph {et~al.}(2024{\natexlab{b}})\citenamefont {Yi}, \citenamefont {You}, \citenamefont {Wu}, \citenamefont {Chen},\ and\ \citenamefont {Liu}}]{Yi:2023npi}%
  \BibitemOpen
  \bibfield  {author} {\bibinfo {author} {\bibfnamefont {Z.}~\bibnamefont {Yi}}, \bibinfo {author} {\bibfnamefont {Z.-Q.}\ \bibnamefont {You}}, \bibinfo {author} {\bibfnamefont {Y.}~\bibnamefont {Wu}}, \bibinfo {author} {\bibfnamefont {Z.-C.}\ \bibnamefont {Chen}}, \ and\ \bibinfo {author} {\bibfnamefont {L.}~\bibnamefont {Liu}},\ }\href {\doibase 10.1088/1475-7516/2024/06/043} {\bibfield  {journal} {\bibinfo  {journal} {JCAP}\ }\textbf {\bibinfo {volume} {06}},\ \bibinfo {pages} {043} (\bibinfo {year} {2024}{\natexlab{b}})},\ \Eprint {http://arxiv.org/abs/2308.14688} {arXiv:2308.14688 [astro-ph.CO]} \BibitemShut {NoStop}%
\bibitem [{\citenamefont {Harigaya}\ \emph {et~al.}(2023)\citenamefont {Harigaya}, \citenamefont {Inomata},\ and\ \citenamefont {Terada}}]{Harigaya:2023pmw}%
  \BibitemOpen
  \bibfield  {author} {\bibinfo {author} {\bibfnamefont {K.}~\bibnamefont {Harigaya}}, \bibinfo {author} {\bibfnamefont {K.}~\bibnamefont {Inomata}}, \ and\ \bibinfo {author} {\bibfnamefont {T.}~\bibnamefont {Terada}},\ }\href {\doibase 10.1103/PhysRevD.108.123538} {\bibfield  {journal} {\bibinfo  {journal} {Phys. Rev. D}\ }\textbf {\bibinfo {volume} {108}},\ \bibinfo {pages} {123538} (\bibinfo {year} {2023})},\ \Eprint {http://arxiv.org/abs/2309.00228} {arXiv:2309.00228 [astro-ph.CO]} \BibitemShut {NoStop}%
\bibitem [{\citenamefont {Chen}\ \emph {et~al.}(2024)\citenamefont {Chen}, \citenamefont {Li}, \citenamefont {Liu},\ and\ \citenamefont {Yi}}]{Chen:2024fir}%
  \BibitemOpen
  \bibfield  {author} {\bibinfo {author} {\bibfnamefont {Z.-C.}\ \bibnamefont {Chen}}, \bibinfo {author} {\bibfnamefont {J.}~\bibnamefont {Li}}, \bibinfo {author} {\bibfnamefont {L.}~\bibnamefont {Liu}}, \ and\ \bibinfo {author} {\bibfnamefont {Z.}~\bibnamefont {Yi}},\ }\href {\doibase 10.1103/PhysRevD.109.L101302} {\bibfield  {journal} {\bibinfo  {journal} {Phys. Rev. D}\ }\textbf {\bibinfo {volume} {109}},\ \bibinfo {pages} {L101302} (\bibinfo {year} {2024})},\ \Eprint {http://arxiv.org/abs/2401.09818} {arXiv:2401.09818 [gr-qc]} \BibitemShut {NoStop}%
\bibitem [{\citenamefont {Chen}\ and\ \citenamefont {Liu}(2024)}]{Chen:2024twp}%
  \BibitemOpen
  \bibfield  {author} {\bibinfo {author} {\bibfnamefont {Z.-C.}\ \bibnamefont {Chen}}\ and\ \bibinfo {author} {\bibfnamefont {L.}~\bibnamefont {Liu}},\ }\href@noop {} {\  (\bibinfo {year} {2024})},\ \Eprint {http://arxiv.org/abs/2402.16781} {arXiv:2402.16781 [astro-ph.CO]} \BibitemShut {NoStop}%
\bibitem [{\citenamefont {Kinney}(1997)}]{Kinney:1997ne}%
  \BibitemOpen
  \bibfield  {author} {\bibinfo {author} {\bibfnamefont {W.~H.}\ \bibnamefont {Kinney}},\ }\href {\doibase 10.1103/PhysRevD.56.2002} {\bibfield  {journal} {\bibinfo  {journal} {Phys. Rev. D}\ }\textbf {\bibinfo {volume} {56}},\ \bibinfo {pages} {2002} (\bibinfo {year} {1997})},\ \Eprint {http://arxiv.org/abs/hep-ph/9702427} {arXiv:hep-ph/9702427} \BibitemShut {NoStop}%
\bibitem [{\citenamefont {Inoue}\ and\ \citenamefont {Yokoyama}(2002)}]{Inoue:2001zt}%
  \BibitemOpen
  \bibfield  {author} {\bibinfo {author} {\bibfnamefont {S.}~\bibnamefont {Inoue}}\ and\ \bibinfo {author} {\bibfnamefont {J.}~\bibnamefont {Yokoyama}},\ }\href {\doibase 10.1016/S0370-2693(01)01369-7} {\bibfield  {journal} {\bibinfo  {journal} {Phys. Lett. B}\ }\textbf {\bibinfo {volume} {524}},\ \bibinfo {pages} {15} (\bibinfo {year} {2002})},\ \Eprint {http://arxiv.org/abs/hep-ph/0104083} {arXiv:hep-ph/0104083} \BibitemShut {NoStop}%
\bibitem [{\citenamefont {Kinney}(2005)}]{Kinney:2005vj}%
  \BibitemOpen
  \bibfield  {author} {\bibinfo {author} {\bibfnamefont {W.~H.}\ \bibnamefont {Kinney}},\ }\href {\doibase 10.1103/PhysRevD.72.023515} {\bibfield  {journal} {\bibinfo  {journal} {Phys. Rev. D}\ }\textbf {\bibinfo {volume} {72}},\ \bibinfo {pages} {023515} (\bibinfo {year} {2005})},\ \Eprint {http://arxiv.org/abs/gr-qc/0503017} {arXiv:gr-qc/0503017} \BibitemShut {NoStop}%
\bibitem [{\citenamefont {Kamenshchik}\ \emph {et~al.}(2019)\citenamefont {Kamenshchik}, \citenamefont {Tronconi}, \citenamefont {Vardanyan},\ and\ \citenamefont {Venturi}}]{Kamenshchik:2018sig}%
  \BibitemOpen
  \bibfield  {author} {\bibinfo {author} {\bibfnamefont {A.~Y.}\ \bibnamefont {Kamenshchik}}, \bibinfo {author} {\bibfnamefont {A.}~\bibnamefont {Tronconi}}, \bibinfo {author} {\bibfnamefont {T.}~\bibnamefont {Vardanyan}}, \ and\ \bibinfo {author} {\bibfnamefont {G.}~\bibnamefont {Venturi}},\ }\href {\doibase 10.1016/j.physletb.2019.02.036} {\bibfield  {journal} {\bibinfo  {journal} {Phys. Lett. B}\ }\textbf {\bibinfo {volume} {791}},\ \bibinfo {pages} {201} (\bibinfo {year} {2019})},\ \Eprint {http://arxiv.org/abs/1812.02547} {arXiv:1812.02547 [gr-qc]} \BibitemShut {NoStop}%
\bibitem [{\citenamefont {Ballesteros}\ \emph {et~al.}(2019)\citenamefont {Ballesteros}, \citenamefont {Beltran~Jimenez},\ and\ \citenamefont {Pieroni}}]{Ballesteros:2018wlw}%
  \BibitemOpen
  \bibfield  {author} {\bibinfo {author} {\bibfnamefont {G.}~\bibnamefont {Ballesteros}}, \bibinfo {author} {\bibfnamefont {J.}~\bibnamefont {Beltran~Jimenez}}, \ and\ \bibinfo {author} {\bibfnamefont {M.}~\bibnamefont {Pieroni}},\ }\href {\doibase 10.1088/1475-7516/2019/06/016} {\bibfield  {journal} {\bibinfo  {journal} {JCAP}\ }\textbf {\bibinfo {volume} {06}},\ \bibinfo {pages} {016} (\bibinfo {year} {2019})},\ \Eprint {http://arxiv.org/abs/1811.03065} {arXiv:1811.03065 [astro-ph.CO]} \BibitemShut {NoStop}%
\bibitem [{\citenamefont {Cai}\ \emph {et~al.}(2018)\citenamefont {Cai}, \citenamefont {Tong}, \citenamefont {Wang},\ and\ \citenamefont {Yan}}]{Cai:2018tuh}%
  \BibitemOpen
  \bibfield  {author} {\bibinfo {author} {\bibfnamefont {Y.-F.}\ \bibnamefont {Cai}}, \bibinfo {author} {\bibfnamefont {X.}~\bibnamefont {Tong}}, \bibinfo {author} {\bibfnamefont {D.-G.}\ \bibnamefont {Wang}}, \ and\ \bibinfo {author} {\bibfnamefont {S.-F.}\ \bibnamefont {Yan}},\ }\href {\doibase 10.1103/PhysRevLett.121.081306} {\bibfield  {journal} {\bibinfo  {journal} {Phys. Rev. Lett.}\ }\textbf {\bibinfo {volume} {121}},\ \bibinfo {pages} {081306} (\bibinfo {year} {2018})},\ \Eprint {http://arxiv.org/abs/1805.03639} {arXiv:1805.03639 [astro-ph.CO]} \BibitemShut {NoStop}%
\bibitem [{\citenamefont {Chen}\ and\ \citenamefont {Cai}(2019)}]{Chen:2019zza}%
  \BibitemOpen
  \bibfield  {author} {\bibinfo {author} {\bibfnamefont {C.}~\bibnamefont {Chen}}\ and\ \bibinfo {author} {\bibfnamefont {Y.-F.}\ \bibnamefont {Cai}},\ }\href {\doibase 10.1088/1475-7516/2019/10/068} {\bibfield  {journal} {\bibinfo  {journal} {JCAP}\ }\textbf {\bibinfo {volume} {10}},\ \bibinfo {pages} {068} (\bibinfo {year} {2019})},\ \Eprint {http://arxiv.org/abs/1908.03942} {arXiv:1908.03942 [astro-ph.CO]} \BibitemShut {NoStop}%
\bibitem [{\citenamefont {Zhou}\ \emph {et~al.}(2020)\citenamefont {Zhou}, \citenamefont {Jiang}, \citenamefont {Cai}, \citenamefont {Sasaki},\ and\ \citenamefont {Pi}}]{Zhou:2020kkf}%
  \BibitemOpen
  \bibfield  {author} {\bibinfo {author} {\bibfnamefont {Z.}~\bibnamefont {Zhou}}, \bibinfo {author} {\bibfnamefont {J.}~\bibnamefont {Jiang}}, \bibinfo {author} {\bibfnamefont {Y.-F.}\ \bibnamefont {Cai}}, \bibinfo {author} {\bibfnamefont {M.}~\bibnamefont {Sasaki}}, \ and\ \bibinfo {author} {\bibfnamefont {S.}~\bibnamefont {Pi}},\ }\href {\doibase 10.1103/PhysRevD.102.103527} {\bibfield  {journal} {\bibinfo  {journal} {Phys. Rev. D}\ }\textbf {\bibinfo {volume} {102}},\ \bibinfo {pages} {103527} (\bibinfo {year} {2020})},\ \Eprint {http://arxiv.org/abs/2010.03537} {arXiv:2010.03537 [astro-ph.CO]} \BibitemShut {NoStop}%
\bibitem [{\citenamefont {Fu}\ \emph {et~al.}(2019)\citenamefont {Fu}, \citenamefont {Wu},\ and\ \citenamefont {Yu}}]{Fu:2019ttf}%
  \BibitemOpen
  \bibfield  {author} {\bibinfo {author} {\bibfnamefont {C.}~\bibnamefont {Fu}}, \bibinfo {author} {\bibfnamefont {P.}~\bibnamefont {Wu}}, \ and\ \bibinfo {author} {\bibfnamefont {H.}~\bibnamefont {Yu}},\ }\href {\doibase 10.1103/PhysRevD.100.063532} {\bibfield  {journal} {\bibinfo  {journal} {Phys. Rev. D}\ }\textbf {\bibinfo {volume} {100}},\ \bibinfo {pages} {063532} (\bibinfo {year} {2019})},\ \Eprint {http://arxiv.org/abs/1907.05042} {arXiv:1907.05042 [astro-ph.CO]} \BibitemShut {NoStop}%
\bibitem [{\citenamefont {Fu}\ \emph {et~al.}(2020)\citenamefont {Fu}, \citenamefont {Wu},\ and\ \citenamefont {Yu}}]{Fu:2019vqc}%
  \BibitemOpen
  \bibfield  {author} {\bibinfo {author} {\bibfnamefont {C.}~\bibnamefont {Fu}}, \bibinfo {author} {\bibfnamefont {P.}~\bibnamefont {Wu}}, \ and\ \bibinfo {author} {\bibfnamefont {H.}~\bibnamefont {Yu}},\ }\href {\doibase 10.1103/PhysRevD.101.023529} {\bibfield  {journal} {\bibinfo  {journal} {Phys. Rev. D}\ }\textbf {\bibinfo {volume} {101}},\ \bibinfo {pages} {023529} (\bibinfo {year} {2020})},\ \Eprint {http://arxiv.org/abs/1912.05927} {arXiv:1912.05927 [astro-ph.CO]} \BibitemShut {NoStop}%
\bibitem [{\citenamefont {Amendola}(1993)}]{Amendola:1993uh}%
  \BibitemOpen
  \bibfield  {author} {\bibinfo {author} {\bibfnamefont {L.}~\bibnamefont {Amendola}},\ }\href {\doibase 10.1016/0370-2693(93)90685-B} {\bibfield  {journal} {\bibinfo  {journal} {Phys. Lett. B}\ }\textbf {\bibinfo {volume} {301}},\ \bibinfo {pages} {175} (\bibinfo {year} {1993})},\ \Eprint {http://arxiv.org/abs/gr-qc/9302010} {arXiv:gr-qc/9302010} \BibitemShut {NoStop}%
\bibitem [{\citenamefont {Kaloper}(2004)}]{Kaloper:2003yf}%
  \BibitemOpen
  \bibfield  {author} {\bibinfo {author} {\bibfnamefont {N.}~\bibnamefont {Kaloper}},\ }\href {\doibase 10.1016/j.physletb.2004.01.005} {\bibfield  {journal} {\bibinfo  {journal} {Phys. Lett. B}\ }\textbf {\bibinfo {volume} {583}},\ \bibinfo {pages} {1} (\bibinfo {year} {2004})},\ \Eprint {http://arxiv.org/abs/hep-ph/0312002} {arXiv:hep-ph/0312002} \BibitemShut {NoStop}%
\bibitem [{\citenamefont {Sushkov}(2009)}]{Sushkov:2009hk}%
  \BibitemOpen
  \bibfield  {author} {\bibinfo {author} {\bibfnamefont {S.~V.}\ \bibnamefont {Sushkov}},\ }\href {\doibase 10.1103/PhysRevD.80.103505} {\bibfield  {journal} {\bibinfo  {journal} {Phys. Rev. D}\ }\textbf {\bibinfo {volume} {80}},\ \bibinfo {pages} {103505} (\bibinfo {year} {2009})},\ \Eprint {http://arxiv.org/abs/0910.0980} {arXiv:0910.0980 [gr-qc]} \BibitemShut {NoStop}%
\bibitem [{\citenamefont {Germani}\ and\ \citenamefont {Kehagias}(2010)}]{Germani:2010gm}%
  \BibitemOpen
  \bibfield  {author} {\bibinfo {author} {\bibfnamefont {C.}~\bibnamefont {Germani}}\ and\ \bibinfo {author} {\bibfnamefont {A.}~\bibnamefont {Kehagias}},\ }\href {\doibase 10.1103/PhysRevLett.105.011302} {\bibfield  {journal} {\bibinfo  {journal} {Phys. Rev. Lett.}\ }\textbf {\bibinfo {volume} {105}},\ \bibinfo {pages} {011302} (\bibinfo {year} {2010})},\ \Eprint {http://arxiv.org/abs/1003.2635} {arXiv:1003.2635 [hep-ph]} \BibitemShut {NoStop}%
\bibitem [{\citenamefont {Germani}\ and\ \citenamefont {Watanabe}(2011)}]{Germani:2011ua}%
  \BibitemOpen
  \bibfield  {author} {\bibinfo {author} {\bibfnamefont {C.}~\bibnamefont {Germani}}\ and\ \bibinfo {author} {\bibfnamefont {Y.}~\bibnamefont {Watanabe}},\ }\href {\doibase 10.1088/1475-7516/2011/07/031} {\bibfield  {journal} {\bibinfo  {journal} {JCAP}\ }\textbf {\bibinfo {volume} {07}},\ \bibinfo {pages} {031} (\bibinfo {year} {2011})},\ \bibinfo {note} {[Addendum: JCAP 07, A01 (2011)]},\ \Eprint {http://arxiv.org/abs/1106.0502} {arXiv:1106.0502 [astro-ph.CO]} \BibitemShut {NoStop}%
\bibitem [{\citenamefont {Tsujikawa}(2012)}]{Tsujikawa:2012mk}%
  \BibitemOpen
  \bibfield  {author} {\bibinfo {author} {\bibfnamefont {S.}~\bibnamefont {Tsujikawa}},\ }\href {\doibase 10.1103/PhysRevD.85.083518} {\bibfield  {journal} {\bibinfo  {journal} {Phys. Rev. D}\ }\textbf {\bibinfo {volume} {85}},\ \bibinfo {pages} {083518} (\bibinfo {year} {2012})},\ \Eprint {http://arxiv.org/abs/1201.5926} {arXiv:1201.5926 [astro-ph.CO]} \BibitemShut {NoStop}%
\bibitem [{\citenamefont {Gialamas}\ \emph {et~al.}(2020)\citenamefont {Gialamas}, \citenamefont {Karam}, \citenamefont {Lykkas},\ and\ \citenamefont {Pappas}}]{Gialamas:2020vto}%
  \BibitemOpen
  \bibfield  {author} {\bibinfo {author} {\bibfnamefont {I.~D.}\ \bibnamefont {Gialamas}}, \bibinfo {author} {\bibfnamefont {A.}~\bibnamefont {Karam}}, \bibinfo {author} {\bibfnamefont {A.}~\bibnamefont {Lykkas}}, \ and\ \bibinfo {author} {\bibfnamefont {T.~D.}\ \bibnamefont {Pappas}},\ }\href {\doibase 10.1103/PhysRevD.102.063522} {\bibfield  {journal} {\bibinfo  {journal} {Phys. Rev. D}\ }\textbf {\bibinfo {volume} {102}},\ \bibinfo {pages} {063522} (\bibinfo {year} {2020})},\ \Eprint {http://arxiv.org/abs/2008.06371} {arXiv:2008.06371 [gr-qc]} \BibitemShut {NoStop}%
\bibitem [{\citenamefont {Gialamas}\ \emph {et~al.}(2024)\citenamefont {Gialamas}, \citenamefont {Katsoulas},\ and\ \citenamefont {Tamvakis}}]{Gialamas:2024jeb}%
  \BibitemOpen
  \bibfield  {author} {\bibinfo {author} {\bibfnamefont {I.~D.}\ \bibnamefont {Gialamas}}, \bibinfo {author} {\bibfnamefont {T.}~\bibnamefont {Katsoulas}}, \ and\ \bibinfo {author} {\bibfnamefont {K.}~\bibnamefont {Tamvakis}},\ }\href {\doibase 10.1088/1475-7516/2024/06/005} {\bibfield  {journal} {\bibinfo  {journal} {JCAP}\ }\textbf {\bibinfo {volume} {06}},\ \bibinfo {pages} {005} (\bibinfo {year} {2024})},\ \Eprint {http://arxiv.org/abs/2403.08530} {arXiv:2403.08530 [gr-qc]} \BibitemShut {NoStop}%
\bibitem [{\citenamefont {Silverstein}\ and\ \citenamefont {Westphal}(2008)}]{Silverstein:2008sg}%
  \BibitemOpen
  \bibfield  {author} {\bibinfo {author} {\bibfnamefont {E.}~\bibnamefont {Silverstein}}\ and\ \bibinfo {author} {\bibfnamefont {A.}~\bibnamefont {Westphal}},\ }\href {\doibase 10.1103/PhysRevD.78.106003} {\bibfield  {journal} {\bibinfo  {journal} {Phys. Rev. D}\ }\textbf {\bibinfo {volume} {78}},\ \bibinfo {pages} {106003} (\bibinfo {year} {2008})},\ \Eprint {http://arxiv.org/abs/0803.3085} {arXiv:0803.3085 [hep-th]} \BibitemShut {NoStop}%
\bibitem [{\citenamefont {Mukhanov}(2005)}]{Mukhanov:2005sc}%
  \BibitemOpen
  \bibfield  {author} {\bibinfo {author} {\bibfnamefont {V.}~\bibnamefont {Mukhanov}},\ }\href {\doibase 10.1017/CBO9780511790553} {\emph {\bibinfo {title} {{Physical Foundations of Cosmology}}}}\ (\bibinfo  {publisher} {Cambridge University Press},\ \bibinfo {address} {Oxford},\ \bibinfo {year} {2005})\BibitemShut {NoStop}%
\bibitem [{\citenamefont {Inomata}\ and\ \citenamefont {Nakama}(2019)}]{Inomata:2018epa}%
  \BibitemOpen
  \bibfield  {author} {\bibinfo {author} {\bibfnamefont {K.}~\bibnamefont {Inomata}}\ and\ \bibinfo {author} {\bibfnamefont {T.}~\bibnamefont {Nakama}},\ }\href {\doibase 10.1103/PhysRevD.99.043511} {\bibfield  {journal} {\bibinfo  {journal} {Phys. Rev. D}\ }\textbf {\bibinfo {volume} {99}},\ \bibinfo {pages} {043511} (\bibinfo {year} {2019})},\ \Eprint {http://arxiv.org/abs/1812.00674} {arXiv:1812.00674 [astro-ph.CO]} \BibitemShut {NoStop}%
\bibitem [{\citenamefont {Liu}\ \emph {et~al.}(2020)\citenamefont {Liu}, \citenamefont {Guo},\ and\ \citenamefont {Cai}}]{Liu:2020oqe}%
  \BibitemOpen
  \bibfield  {author} {\bibinfo {author} {\bibfnamefont {J.}~\bibnamefont {Liu}}, \bibinfo {author} {\bibfnamefont {Z.-K.}\ \bibnamefont {Guo}}, \ and\ \bibinfo {author} {\bibfnamefont {R.-G.}\ \bibnamefont {Cai}},\ }\href {\doibase 10.1103/PhysRevD.101.083535} {\bibfield  {journal} {\bibinfo  {journal} {Phys. Rev. D}\ }\textbf {\bibinfo {volume} {101}},\ \bibinfo {pages} {083535} (\bibinfo {year} {2020})},\ \Eprint {http://arxiv.org/abs/2003.02075} {arXiv:2003.02075 [astro-ph.CO]} \BibitemShut {NoStop}%
\bibitem [{\citenamefont {Kerr}\ \emph {et~al.}(2020)\citenamefont {Kerr} \emph {et~al.}}]{Kerr:2020qdo}%
  \BibitemOpen
  \bibfield  {author} {\bibinfo {author} {\bibfnamefont {M.}~\bibnamefont {Kerr}} \emph {et~al.},\ }\href {\doibase 10.1017/pasa.2020.11} {\bibfield  {journal} {\bibinfo  {journal} {Publ. Astron. Soc. Austral.}\ }\textbf {\bibinfo {volume} {37}},\ \bibinfo {pages} {e020} (\bibinfo {year} {2020})},\ \Eprint {http://arxiv.org/abs/2003.09780} {arXiv:2003.09780 [astro-ph.IM]} \BibitemShut {NoStop}%
\bibitem [{\citenamefont {Goncharov}\ \emph {et~al.}(2021)\citenamefont {Goncharov} \emph {et~al.}}]{Goncharov:2020krd}%
  \BibitemOpen
  \bibfield  {author} {\bibinfo {author} {\bibfnamefont {B.}~\bibnamefont {Goncharov}} \emph {et~al.},\ }\href {\doibase 10.1093/mnras/staa3411} {\bibfield  {journal} {\bibinfo  {journal} {Mon. Not. Roy. Astron. Soc.}\ }\textbf {\bibinfo {volume} {502}},\ \bibinfo {pages} {478} (\bibinfo {year} {2021})},\ \Eprint {http://arxiv.org/abs/2010.06109} {arXiv:2010.06109 [astro-ph.HE]} \BibitemShut {NoStop}%
\bibitem [{\citenamefont {Reardon}\ \emph {et~al.}(2023{\natexlab{b}})\citenamefont {Reardon} \emph {et~al.}}]{Reardon:2023zen}%
  \BibitemOpen
  \bibfield  {author} {\bibinfo {author} {\bibfnamefont {D.~J.}\ \bibnamefont {Reardon}} \emph {et~al.},\ }\href {\doibase 10.3847/2041-8213/acdd03} {\bibfield  {journal} {\bibinfo  {journal} {Astrophys. J. Lett.}\ }\textbf {\bibinfo {volume} {951}},\ \bibinfo {pages} {L7} (\bibinfo {year} {2023}{\natexlab{b}})},\ \Eprint {http://arxiv.org/abs/2306.16229} {arXiv:2306.16229 [astro-ph.HE]} \BibitemShut {NoStop}%
\bibitem [{\citenamefont {Arzoumanian}\ \emph {et~al.}(2016)\citenamefont {Arzoumanian} \emph {et~al.}}]{NANOGrav:2015aud}%
  \BibitemOpen
  \bibfield  {author} {\bibinfo {author} {\bibfnamefont {Z.}~\bibnamefont {Arzoumanian}} \emph {et~al.} (\bibinfo {collaboration} {NANOGrav}),\ }\href {\doibase 10.3847/0004-637X/821/1/13} {\bibfield  {journal} {\bibinfo  {journal} {Astrophys. J.}\ }\textbf {\bibinfo {volume} {821}},\ \bibinfo {pages} {13} (\bibinfo {year} {2016})},\ \Eprint {http://arxiv.org/abs/1508.03024} {arXiv:1508.03024 [astro-ph.GA]} \BibitemShut {NoStop}%
\bibitem [{\citenamefont {Ellis}\ \emph {et~al.}(2020)\citenamefont {Ellis}, \citenamefont {Vallisneri}, \citenamefont {Taylor},\ and\ \citenamefont {Baker}}]{enterprise}%
  \BibitemOpen
  \bibfield  {author} {\bibinfo {author} {\bibfnamefont {J.~A.}\ \bibnamefont {Ellis}}, \bibinfo {author} {\bibfnamefont {M.}~\bibnamefont {Vallisneri}}, \bibinfo {author} {\bibfnamefont {S.~R.}\ \bibnamefont {Taylor}}, \ and\ \bibinfo {author} {\bibfnamefont {P.~T.}\ \bibnamefont {Baker}},\ }\href {\doibase 10.5281/zenodo.4059815} {\enquote {\bibinfo {title} {Enterprise: Enhanced numerical toolbox enabling a robust pulsar inference suite},}\ }\bibinfo {howpublished} {Zenodo} (\bibinfo {year} {2020})\BibitemShut {NoStop}%
\bibitem [{\citenamefont {Thrane}\ and\ \citenamefont {Romano}(2013)}]{Thrane:2013oya}%
  \BibitemOpen
  \bibfield  {author} {\bibinfo {author} {\bibfnamefont {E.}~\bibnamefont {Thrane}}\ and\ \bibinfo {author} {\bibfnamefont {J.~D.}\ \bibnamefont {Romano}},\ }\href {\doibase 10.1103/PhysRevD.88.124032} {\bibfield  {journal} {\bibinfo  {journal} {Phys. Rev. D}\ }\textbf {\bibinfo {volume} {88}},\ \bibinfo {pages} {124032} (\bibinfo {year} {2013})},\ \Eprint {http://arxiv.org/abs/1310.5300} {arXiv:1310.5300 [astro-ph.IM]} \BibitemShut {NoStop}%
\bibitem [{\citenamefont {Aghanim}\ \emph {et~al.}(2020)\citenamefont {Aghanim} \emph {et~al.}}]{Planck:2018vyg}%
  \BibitemOpen
  \bibfield  {author} {\bibinfo {author} {\bibfnamefont {N.}~\bibnamefont {Aghanim}} \emph {et~al.} (\bibinfo {collaboration} {Planck}),\ }\href {\doibase 10.1051/0004-6361/201833910} {\bibfield  {journal} {\bibinfo  {journal} {Astron. Astrophys.}\ }\textbf {\bibinfo {volume} {641}},\ \bibinfo {pages} {A6} (\bibinfo {year} {2020})},\ \bibinfo {note} {[Erratum: Astron.Astrophys. 652, C4 (2021)]},\ \Eprint {http://arxiv.org/abs/1807.06209} {arXiv:1807.06209 [astro-ph.CO]} \BibitemShut {NoStop}%
\bibitem [{\citenamefont {Kass}\ and\ \citenamefont {Raftery}(1995)}]{BF}%
  \BibitemOpen
  \bibfield  {author} {\bibinfo {author} {\bibfnamefont {R.~E.}\ \bibnamefont {Kass}}\ and\ \bibinfo {author} {\bibfnamefont {A.~E.}\ \bibnamefont {Raftery}},\ }\href {\doibase 10.1080/01621459.1995.10476572} {\bibfield  {journal} {\bibinfo  {journal} {Journal of the American Statistical Association}\ }\textbf {\bibinfo {volume} {90}},\ \bibinfo {pages} {773} (\bibinfo {year} {1995})}\BibitemShut {NoStop}%
\bibitem [{\citenamefont {Taylor}\ \emph {et~al.}(2021)\citenamefont {Taylor}, \citenamefont {Baker}, \citenamefont {Hazboun}, \citenamefont {Simon},\ and\ \citenamefont {Vigeland}}]{enterprise_ext}%
  \BibitemOpen
  \bibfield  {author} {\bibinfo {author} {\bibfnamefont {S.~R.}\ \bibnamefont {Taylor}}, \bibinfo {author} {\bibfnamefont {P.~T.}\ \bibnamefont {Baker}}, \bibinfo {author} {\bibfnamefont {J.~S.}\ \bibnamefont {Hazboun}}, \bibinfo {author} {\bibfnamefont {J.}~\bibnamefont {Simon}}, \ and\ \bibinfo {author} {\bibfnamefont {S.~J.}\ \bibnamefont {Vigeland}},\ }\href {https://github.com/nanograv/enterprise_extensions} {\enquote {\bibinfo {title} {enterprise\_extensions},}\ } (\bibinfo {year} {2021}),\ \bibinfo {note} {v2.4.3}\BibitemShut {NoStop}%
\bibitem [{\citenamefont {Mitridate}\ \emph {et~al.}(2023)\citenamefont {Mitridate}, \citenamefont {Wright}, \citenamefont {von Eckardstein}, \citenamefont {Schr\"oder}, \citenamefont {Nay}, \citenamefont {Olum}, \citenamefont {Schmitz},\ and\ \citenamefont {Trickle}}]{Mitridate:2023oar}%
  \BibitemOpen
  \bibfield  {author} {\bibinfo {author} {\bibfnamefont {A.}~\bibnamefont {Mitridate}}, \bibinfo {author} {\bibfnamefont {D.}~\bibnamefont {Wright}}, \bibinfo {author} {\bibfnamefont {R.}~\bibnamefont {von Eckardstein}}, \bibinfo {author} {\bibfnamefont {T.}~\bibnamefont {Schr\"oder}}, \bibinfo {author} {\bibfnamefont {J.}~\bibnamefont {Nay}}, \bibinfo {author} {\bibfnamefont {K.}~\bibnamefont {Olum}}, \bibinfo {author} {\bibfnamefont {K.}~\bibnamefont {Schmitz}}, \ and\ \bibinfo {author} {\bibfnamefont {T.}~\bibnamefont {Trickle}},\ }\href@noop {} {\  (\bibinfo {year} {2023})},\ \Eprint {http://arxiv.org/abs/2306.16377} {arXiv:2306.16377 [hep-ph]} \BibitemShut {NoStop}%
\bibitem [{\citenamefont {Mitridate}(2023)}]{andrea_mitridate_2023}%
  \BibitemOpen
  \bibfield  {author} {\bibinfo {author} {\bibfnamefont {A.}~\bibnamefont {Mitridate}},\ }\href {\doibase 10.5281/zenodo.7876430} {\  (\bibinfo {year} {2023}),\ 10.5281/zenodo.7876430}\BibitemShut {NoStop}%
\bibitem [{\citenamefont {Carlin}\ and\ \citenamefont {Chib}(1995)}]{10.2307/2346151}%
  \BibitemOpen
  \bibfield  {author} {\bibinfo {author} {\bibfnamefont {B.~P.}\ \bibnamefont {Carlin}}\ and\ \bibinfo {author} {\bibfnamefont {S.}~\bibnamefont {Chib}},\ }\href {http://www.jstor.org/stable/2346151} {\bibfield  {journal} {\bibinfo  {journal} {Journal of the Royal Statistical Society. Series B (Methodological)}\ }\textbf {\bibinfo {volume} {57}},\ \bibinfo {pages} {473} (\bibinfo {year} {1995})}\BibitemShut {NoStop}%
\bibitem [{\citenamefont {Godsill}(2001)}]{10.2307/1391010}%
  \BibitemOpen
  \bibfield  {author} {\bibinfo {author} {\bibfnamefont {S.~J.}\ \bibnamefont {Godsill}},\ }\href {http://www.jstor.org/stable/1391010} {\bibfield  {journal} {\bibinfo  {journal} {Journal of Computational and Graphical Statistics}\ }\textbf {\bibinfo {volume} {10}},\ \bibinfo {pages} {230} (\bibinfo {year} {2001})}\BibitemShut {NoStop}%
\bibitem [{\citenamefont {Hee}\ \emph {et~al.}(2016)\citenamefont {Hee}, \citenamefont {Handley}, \citenamefont {Hobson},\ and\ \citenamefont {Lasenby}}]{Hee:2015eba}%
  \BibitemOpen
  \bibfield  {author} {\bibinfo {author} {\bibfnamefont {S.}~\bibnamefont {Hee}}, \bibinfo {author} {\bibfnamefont {W.}~\bibnamefont {Handley}}, \bibinfo {author} {\bibfnamefont {M.~P.}\ \bibnamefont {Hobson}}, \ and\ \bibinfo {author} {\bibfnamefont {A.~N.}\ \bibnamefont {Lasenby}},\ }\href {\doibase 10.1093/mnras/stv2217} {\bibfield  {journal} {\bibinfo  {journal} {Mon. Not. Roy. Astron. Soc.}\ }\textbf {\bibinfo {volume} {455}},\ \bibinfo {pages} {2461} (\bibinfo {year} {2016})},\ \Eprint {http://arxiv.org/abs/1506.09024} {arXiv:1506.09024 [astro-ph.CO]} \BibitemShut {NoStop}%
\bibitem [{\citenamefont {Taylor}\ \emph {et~al.}(2020)\citenamefont {Taylor}, \citenamefont {van Haasteren},\ and\ \citenamefont {Sesana}}]{Taylor:2020zpk}%
  \BibitemOpen
  \bibfield  {author} {\bibinfo {author} {\bibfnamefont {S.~R.}\ \bibnamefont {Taylor}}, \bibinfo {author} {\bibfnamefont {R.}~\bibnamefont {van Haasteren}}, \ and\ \bibinfo {author} {\bibfnamefont {A.}~\bibnamefont {Sesana}},\ }\href {\doibase 10.1103/PhysRevD.102.084039} {\bibfield  {journal} {\bibinfo  {journal} {Phys. Rev. D}\ }\textbf {\bibinfo {volume} {102}},\ \bibinfo {pages} {084039} (\bibinfo {year} {2020})},\ \Eprint {http://arxiv.org/abs/2006.04810} {arXiv:2006.04810 [astro-ph.IM]} \BibitemShut {NoStop}%
\bibitem [{\citenamefont {Kobayashi}\ \emph {et~al.}(2011)\citenamefont {Kobayashi}, \citenamefont {Yamaguchi},\ and\ \citenamefont {Yokoyama}}]{Kobayashi:2011nu}%
  \BibitemOpen
  \bibfield  {author} {\bibinfo {author} {\bibfnamefont {T.}~\bibnamefont {Kobayashi}}, \bibinfo {author} {\bibfnamefont {M.}~\bibnamefont {Yamaguchi}}, \ and\ \bibinfo {author} {\bibfnamefont {J.}~\bibnamefont {Yokoyama}},\ }\href {\doibase 10.1143/PTP.126.511} {\bibfield  {journal} {\bibinfo  {journal} {Prog. Theor. Phys.}\ }\textbf {\bibinfo {volume} {126}},\ \bibinfo {pages} {511} (\bibinfo {year} {2011})},\ \Eprint {http://arxiv.org/abs/1105.5723} {arXiv:1105.5723 [hep-th]} \BibitemShut {NoStop}%
\end{thebibliography}%

\clearpage
\newpage
\maketitle
\onecolumngrid
\begin{center}
\textbf{\large \papertitle} 
\\ 
\vspace{0.05in}
{Chang Han, Li-Yang Chen, Zu-Cheng Chen, Chengjie Fu, Puxun Wu, Hongwei Yu, N. D. Ramesh Bhat, Xiaojin Liu, Valentina Di Marco, Saurav Mishra, Daniel J. Reardon, Christopher J. Russell, Ryan M. Shannon, Lei Zhang, Xingjiang Zhu, and Andrew Zic}
\\ 
\vspace{0.05in}
{ \it Supplementary Material}
\end{center}

\onecolumngrid
\setcounter{equation}{0}
\setcounter{figure}{0}
\setcounter{section}{0}
\setcounter{table}{0}
\setcounter{page}{1}
\makeatletter
\renewcommand{\theequation}{S\arabic{equation}}
\renewcommand{\thefigure}{S\arabic{figure}}
\renewcommand{\thetable}{S\arabic{table}}

\section{Mukhanov-Sasaki equation}\label{appendixA}

Starting from the action given in Eq.~(1), we can derive the second-order action for the curvature perturbation $\mathcal{R}$~\cite{Mukhanov:2005sc,Tsujikawa:2012mk,Kobayashi:2011nu,Fu:2019ttf}:
\begin{equation}
    S^{(2)}=\int dt d^3xa^2Q_s\left [\dot{\mathcal{R}}^2-\frac{c_s^2}{a^2}(\partial_i\mathcal{R})^2 \right],
\end{equation}
where the kinetic coefficient $Q_s$ is given by
\begin{equation}
    Q_s=\frac{w_1(4w_1w_3+9w^2_2)}{3w^2_2}.
\end{equation}
Here the coefficients $w_i$ are defined as:
\begin{align}
    w_1 &= M_P^2(1-2\delta_D), \\
    w_2 &= 2HM_P^2(1-6\delta_D), \\
    w_3 &= -3H^2M_P^2(3-\delta_X+36\delta_D), \\
    w_4 &= M_P^2(1+2\delta_D)
\end{align}
with
\begin{equation}
    \delta_D = \frac{\kappa^4\dot{\phi}^2\theta}{4}, \quad 
    \delta_X = \frac{\kappa^2\dot{\phi}^2}{2H^2}.
\end{equation}

From this second-order action, we can derive the Mukhanov-Sasaki equation:
\begin{equation}
    \label{eq:MuSa}
    u_k'' + \left(c_s^2k^2-\frac{z''}{z}\right)u_k = 0,
\end{equation}
where we have introduced a variable $u_k\equiv z\mathcal{R}_k$ with $z \equiv a\sqrt{2Q_s}$.  The power spectrum of curvature perturbations can then be expressed as:
\begin{equation}
    \mathcal{P}_\mathcal{R}(k) = \frac{k^3}{2\pi^2}\left|\frac{u_k}{z}\right|^2.
\end{equation}

\section{Approximate Solution for $H/\dot{\phi}$}

In this appendix, we derive the solution presented in Eq.~(10) by analyzing different evolutionary stages of the system. The full dynamics are governed by Eqs.~(3) and (4), which can be simplified based on the studies in~\cite{Fu:2019ttf} showing that $-\dot{H}/H^2 \ll 1$ and $\kappa^2\dot{\phi}^2/(2H^2)\ll 1$ hold throughout both the SR and USR stages. These conditions allow us to write
\begin{equation}\label{eq:HV2}
       3H^2\simeq \kappa^2V,
\end{equation}
\begin{equation}\label{eq:dotphi2}
\left(1+3\kappa^2 H^2\theta\right)\ddot{\phi}+3H\left(1+3\kappa^2 H^2\theta\right)\dot{\phi}+\frac{3}{2}\kappa^2H^2\theta _{,\phi}\dot\phi^2+V_{,\phi}\simeq 0.
\end{equation}
In the following we analyze these equations in three distinct regimes.

\subsection{Slow-Roll Regime}
In the SR regime, we have $\ddot{\phi}/H\dot{\phi}\ll 1$, $-\dot{H}/H^2 \ll 1$, and $\theta_{,\phi}\ll\theta$. Although $\theta$ is small, it still contributes to the number of e-folds $N$. Under these conditions, \Eq{eq:dotphi2} reduces to
\begin{equation}
       3H\left(1+3\kappa^2H^2\theta\right)\dot{\phi}+V_{,\phi}\simeq 0,
\end{equation}
which  yields the solution
\begin{equation}
\label{eq:SRsolution}
    \frac{H}{\dot{\phi}}=\frac{-3H^2(1+3\kappa^2H^2\theta)}{V_{,\phi}}.
\end{equation}

\subsection{Ultra-Slow-Roll Regime}
In the USR regime, where $|\phi-\phi_c|<\sigma/\kappa$, the condition $\kappa^2\dot{\phi}^2/(2H^2)\ll 1$ implies
\begin{equation}
    (1+9\kappa^2 H^2\theta)\dot{\phi}^2\simeq 0.
\end{equation}
Taking the time derivative on the above equation and using $|\dot{H}/H^2|\ll 1$ lead to
\begin{equation}
    \ddot{\phi}=-\frac{1}{2}\frac{9\kappa^2H^2\theta_{,\phi}}{1+9\kappa^2H^2\theta}\dot{\phi}^2.
\end{equation}
Substituting this into \Eq{eq:dotphi2} and noting that $\kappa^2H^2\theta \gg 1$ in this regime, one can obtain
\begin{equation}
\label{eq:USRsolution}
    \frac{H}{\dot{\phi}}=-\frac{9\kappa^2H^4\theta}{V_{,\phi}}.
\end{equation}

\subsection{Transition Regime}
Between the SR and USR stages ($1\gg|\phi-\phi_c|\gg\sigma/\kappa$), there is a transition stage. Previous studies in~\cite{Fu:2019vqc} found a linear solution $\dot{\phi}\propto (\phi-\phi_c)$, implying
\begin{equation}
\label{eq:dotdotphihdotphi}
    \ddot{\phi}=\frac{1}{\phi-\phi_c}\dot{\phi}^2,
\end{equation}
which is approximately equivalent to
\begin{equation}
\label{eq:dotdotphihdotphi+th}
    \ddot{\phi}=-\frac{3\kappa^2H^2\theta_{,\phi}}{1+3\kappa^2H^2\theta}\dot{\phi}^2.
\end{equation}
Substituting this into \Eq{eq:dotphi2} and requiring consistency with the SR solution when $\theta=\theta_{,\phi}=0$, we have
\begin{equation}
\label{eq:transsolution}
     \frac{H}{\dot\phi}=-\frac{3H^2\left(3(1+3\kappa^2H^2\theta)+\sqrt{6V_{,\phi}\theta_{,\phi}+9(1+3\kappa^2H^2\theta)^2}\right)}{6V_{,\phi}}.
\end{equation}
Combining Eqs.~\eqref{eq:SRsolution}, \eqref{eq:transsolution}, and \eqref{eq:USRsolution} with \eqref{eq:HV2} yields Eq.~(10).

\section{Derivation of the mapping between $\phi$ and $k$}\label{appendixC}

Here, we provide a detailed derivation of the mapping relationship between  $\phi$ and $k$ from Eq.~(10). We employ several approximations to make the integration tractable while maintaining physical consistency in different regions of the field space.

\subsection{Basic Mapping for Minimal derivative Coupling}

We first consider the minimal derivative coupling case where $\theta=0$. In this limit, Eq.~(10) reduces to
\begin{equation}\label{eq:zero}
      \frac{H}{\dot\phi}\simeq-\frac{\kappa^2V}{V_{,\phi}}.
\end{equation}
Combining  Eqs.~(5), (12), (3), and (11), we obtain
\begin{equation}
       k(\phi)=\sqrt{\frac{\lambda}{3}}\phi^{\frac{p}{2}}e^{-\frac{\kappa^2\phi^2}{2p}}.
\end{equation}
Since the power-law term $\phi^{p/2}$ varies slowly compared to the exponential term, we can approximate it at a fixed value, e.g., at $\phi_c$, yielding the inverse function:
\begin{equation}
    \phi(k)=\sqrt{-\frac{2p}{\kappa^2} \ln\left(\frac{\sqrt{3}k }{\sqrt{\lambda \phi_c^p} } \right)}.
\end{equation}

\subsection{Extension to Non-minimal Coupling}

For the non-minimal coupling case with $\theta$ given by Eq.~(2), we need to consider different regions of the field space. We find that Eq.~(10) can be expressed detailedly  as:
\begin{equation}\label{C4}
    \frac{H}{\dot{\phi } }\simeq 
\left\{\begin{matrix}
-\frac{\kappa ^2 V(1+\kappa^4V\theta)}{V_{,\phi}}, & \mathrm{for} \ \left | \phi-\phi_c \right | >\kappa ^2\omega \lambda \sigma \phi_c^p
 \\ \\
-\frac{\kappa ^2 V \left ( 3(1+\kappa ^4V\theta)+3+\sqrt{6V_{,\phi}\theta _{,\phi} +9(\kappa ^4V\theta)^2}  \right ) }{6V_{,\phi}}, & \ \ \ \mathrm{for} \  \kappa ^2\omega \lambda \sigma \phi_c^p>\left | \phi-\phi_c \right |>\frac{\sigma}{\kappa}
\\ \\ 
-\frac{\kappa ^6V^2\theta}{V_{,\phi}}, & \mathrm{for} \  \frac{\sigma}{\kappa}>\left | \phi-\phi_c \right |.
\end{matrix}\right.
\end{equation}
The different regions  of $|\phi-\phi_c|$ in Eq.~(\ref{C4}) are chosen based on physical considerations. The first region  $|\phi-\phi_c| >\kappa ^2\omega \lambda \sigma \phi_c^p$  comes from the condition $\kappa^4V\theta <1$. In this regime, we have $-6V_{,\phi}\theta_{,\phi} \ll 9(1+\kappa ^4V\theta)^2$, allowing us to neglect the $\theta_{,\phi}$ term. When $\kappa^4V\theta > 1$, we find  $\theta$ grows rapidly, which means that  $V_{,\phi}\theta_{,\phi}>1$ and $\kappa^4V\theta\gg1$ are satisfied quickly. Then, we have  $\sqrt{6V_{,\phi}\theta_{,\phi}+9(1+\kappa^4V\theta)^2}\simeq 3+\sqrt{6V_{,\phi}\theta_{,\phi}+9(\kappa^4V\theta)^2}$.
Further considering the properties of  coupling function shown in \ref{Sub3}, we divide the region of  $\kappa^4V\theta > 1$ into two parts.

Since $V$ and $V_{,\phi}$ vary much more slowly than $\theta$ and $\theta_{,\phi}$, we can use $V^c=V(\phi_c)$ and  $V^c_{,\phi}= V_{,\phi}(\phi_c)$ to replace $V$ and $V_{,\phi}$ when they appear in terms including $\theta$ and $\theta_{,\phi}$. 
 So, Eq.~(\ref{C4}) can be approximated as
\begin{equation}
\label{eq:h/dotphic}
    \frac{H}{\dot{\phi } }\simeq 
\left\{\begin{matrix}
-\frac{\kappa ^2 V}{V_{,\phi}} -\frac{\kappa ^6{V^c}^2\theta }{V^c_{,\phi}},
 & \mathrm{for} \  \left | \phi-\phi_c \right | >\kappa ^2\omega \lambda \sigma \phi_c^p
 \\ \\
-\frac{\kappa ^2 V}{V_{,\phi}}-\frac{\kappa ^6 {V^c}^2\theta}{2V^c_{,\phi}}
-\frac{\kappa ^2 V^c \left ( \sqrt{6V^c_{,\phi}\theta _{,\phi} +9(\kappa ^4V^c\theta )^2}  \right ) }{6V^c_{,\phi}} ,
 &\ \ \ \mathrm{for} \  \kappa ^2\omega \lambda \sigma \phi_c^p>\left | \phi-\phi_c \right |>\frac{\sigma}{\kappa}
 \\ \\ 
-\frac{\kappa ^6{V^c}^2\theta }{{V^c}_{,\phi}},& \mathrm{for} \ \frac{\sigma}{\kappa}>\left | \phi-\phi_c \right |.
\end{matrix}\right.
\end{equation}

\subsection{Approximation of the Coupling Function}\label{Sub3}

The coupling function $\theta$ can be approximated as~\cite{Fu:2019vqc}
\begin{align}\label{C6}
\theta&\simeq 
    \begin{cases}
        +\frac{\omega \sigma }{\kappa(\phi-\phi_c)}, & \text{for } \phi>\phi_c+\frac{\sigma}{\kappa} \\[2ex]
        \frac{\omega \sigma }{\sqrt{\kappa ^2(\phi-\phi_c)^2+\sigma^2}}, & \text{for } |\phi-\phi_c| \leq \frac{\sigma}{\kappa} \\[2ex]
        -\frac{\omega \sigma }{\kappa(\phi-\phi_c)}, & \text{for } \phi<\phi_c-\frac{\sigma}{\kappa}
    \end{cases} \end{align}
and  $\theta_{,\phi}$ can be expressed as 
\begin{align} \label{C7}
\theta_{,\phi} &\simeq  
    \begin{cases}
        -\frac{\omega \sigma }{\kappa ^2(\phi-\phi_c)^2}, & \text{for } \phi>\phi_c \\[2ex]
        +\frac{\omega \sigma }{\kappa ^2(\phi-\phi_c)^2}, & \text{for } \phi<\phi_c .
    \end{cases}
\end{align}

\subsection{Derivation of the Final Mapping Relation}

Using Eqs.~(\ref{C6}) and (\ref{C7}), we find that  \Eq{eq:h/dotphic} can be further expressed as:
\begin{equation}
\frac{H}{\dot{\phi}} \simeq 
\left\{\begin{array}{ll}
-\frac{\kappa^2V}{V_{,\phi}} -\frac{\kappa^6{V^c}^2}{V^c_{,\phi}}\frac{\omega\sigma}{\kappa(\phi-\phi_c)},
& \phi > \phi_c + \kappa^2\omega\lambda\sigma\phi_c^p
\\ \\
-\frac{\kappa^2V}{V_{,\phi}} + \mathcal{F}_1\frac{\omega\sigma}{\kappa(\phi-\phi_c)},
& \phi_c + \frac{\sigma}{\kappa} < \phi < \phi_c + \kappa^2\omega\lambda\sigma\phi_c^p
\\ \\
-\frac{\kappa^6{V^c}^2}{V^c_{,\phi}}\frac{\omega\sigma}{\sqrt{\kappa^2(\phi-\phi_c)^2 + \sigma^2}},
& |\phi - \phi_c| < \frac{\sigma}{\kappa}
\\ \\
-\frac{\kappa^2V}{V_{,\phi}} - \mathcal{F}_2\frac{\omega\sigma}{\kappa(\phi-\phi_c)},
& \phi_c - \kappa^2\omega\lambda\sigma\phi_c^p < \phi < \phi_c - \frac{\sigma}{\kappa}
\\ \\
-\frac{\kappa^2V}{V_{,\phi}} + \frac{\kappa^6{V^c}^2}{V^c_{,\phi}}\frac{\omega\sigma}{\kappa(\phi-\phi_c)},
& \phi < \phi_c - \kappa^2\omega\lambda\sigma\phi_c^p
\end{array}\right.
\end{equation}
where  
\begin{align}
\mathcal{F}_1 &= -\frac{\kappa^6{V^c}^2}{2V^c_{,\phi}} - 
\frac{\kappa^2V^c}{6V^c_{,\phi}}\sqrt{-6V^c_{,\phi}\frac{\kappa^2}{\omega\sigma} + 9(\kappa^4V^c)^2},\\
\mathcal{F}_2 &= -\frac{\kappa^6{V^c}^2}{2V^c_{,\phi}} - 
\frac{\kappa^2V^c}{6V^c_{,\phi}}\sqrt{6V^c_{,\phi}\frac{\kappa^2}{\omega\sigma} + 9(\kappa^4V^c)^2}.
\end{align}

While these approximations are not exact, they are sufficient for our calculations. Integrating this piecewise function according to Eq.~(11) yields $N(\phi)$ and consequently $k(\phi)$. After appropriate approximations, we can obtain the inverse function $\phi(k)$ (setting $\kappa^{-1} = M_P = 1$ for simplicity):
\begin{equation}
\label{eq:phik}
\phi(k) = 
\begin{cases}
\phi_1(k), & 0< k < k_1 \\
\phi_2(k), & k_1 < k < k_2 \\
\phi_3(k), & k_2 < k < k_3 \\
\phi_4(k), & k_3 < k < k_4 \\
\phi_5(k), & k_4 < k
\end{cases}
\end{equation}
where the transition wavenumbers are given by:
\begin{align}
k_1 &= \left(\frac{\lambda(\phi_c + \omega\lambda\sigma\phi_c^p)^p}{3}\right)^{\frac{1}{2}}{a_\mathrm{end}}e^{N_1 + N_2 + N_3 + N_4}, \\
k_2 &= \left(\frac{\lambda(\phi_c + \sigma)^p}{3}\right)^{\frac{1}{2}}{a_\mathrm{end}}e^{N_1 + N_2 + N_3}, \\
k_3 &= \left(\frac{\lambda(\phi_c - \sigma)^p}{3}\right)^{\frac{1}{2}}{a_\mathrm{end}}e^{N_1 + N_2}, \\
k_4 &= \left(\frac{\lambda(\phi_c - \omega\lambda\sigma\phi_c^p)^p}{3}\right)^{\frac{1}{2}}{a_\mathrm{end}}e^{N_1}.
\end{align}
Here the auxiliary functions $N_i$  are defined as:
\begin{align}
N_1 &= \frac{2\lambda\sigma\phi_c^{1+p}\omega\ln[\lambda\sigma\phi_c^{-1+p}\omega]-(\phi_c - \lambda\sigma\phi_c^p\omega)^2 }{2p}, \\
N_2 &= \frac{(\phi_c - \lambda\sigma\phi_c^p\omega)^2 - (\phi_c-\sigma)^2}{2p} + \mathcal{G}_1\ln\left[\frac{1}{\lambda\omega\phi_c^p}\right], \\
N_3 &= \frac{\lambda\omega\sigma\phi_c^{1+p}}{2p}\ln\left[\frac{(\sqrt{2\sigma}-\sigma)^2}{(\sqrt{2\sigma}+\sigma)^2}\right], \\
N_4 &= \frac{(\phi_c+\sigma)^2-(\phi_c+\lambda\omega\sigma\phi_c^p)^2}{2p} + 
\mathcal{G}_2\ln\left[\frac{1}{\lambda\omega\phi_c^p}\right],
\end{align}
with
\begin{align}
\mathcal{G}_1 &= \frac{\lambda\omega\sigma\phi_c}{6p}\left(3\phi_c^p + \sqrt{3\phi_c^p[3\phi_c^p+2p(\sigma\lambda\omega\phi_c)^{-1}]}\right), \\
\mathcal{G}_2 &= \frac{\lambda\omega\sigma\phi_c}{6p}\left(3\phi_c^p + \sqrt{3\phi_c^p[3\phi_c^p-2p(\sigma\lambda\omega\phi_c)^{-1}]}\right).
\end{align}

The corresponding field values in each region are given by:
\begin{align}
\phi_1(k) &= \left\{2p(N_1+N_2+N_3+N_4) + 2(\phi_c+\lambda\omega\sigma\phi_c^p)^2 - (\phi_c+\sigma)^2 
-2p\ln\left[\frac{\sqrt{3}k}{  \sqrt{\lambda\phi_c^p}{a_\mathrm{end}}}\right]\right\}^{\frac{1}{2}}, \\[2ex]
\phi_2(k) &= \phi_c + \sigma\left(\frac{\sqrt{3}e^{-N_1-N_2-N_3}k}{\sqrt{\lambda\phi_c^p}{a_\mathrm{end}}}\right)^{-3+\alpha}, \\[2ex]
\phi_3(k) &= (\phi_c+\sigma)\left[\left(\frac{\lambda(\phi_c+\sigma)}{3}\right)^{\frac{1}{2}}
{a_\mathrm{end}}e^{N_1+N_2+N_3}\right]^{-\beta}k^{\beta}, \\[2ex]
\phi_4(k) &= \phi_c - \sigma\lambda\omega\phi_c^p\left(\frac{\sqrt{3}e^{-N_1+\gamma}k}{\sqrt{\lambda\phi_c^p}{a_\mathrm{end}}}\right)^{-3+\delta}, \\[2ex]
\phi_5(k) &= \left\{2p(N_1+N_2) + (\phi_c-\omega\lambda\sigma\phi_c^p)^2 - 2p\ln\left[\frac{\sqrt{3}k}{\sqrt{\lambda\phi_c^p}{a_\mathrm{end}}}\right] + \epsilon\right\}^{\frac{1}{2}},
\end{align}
where  
\begin{align}
\alpha &= \sqrt{3}\lambda^{-1}\phi_c^{-p}\sqrt{\lambda\phi_c^p(3\lambda\phi_c^p-2p(\sigma\omega\phi_c)^{-1})}, \\
\beta &= N_3^{-1}\ln\left(\frac{\phi_c+\sigma}{\phi_c-\sigma}\right), \\
\gamma &= \frac{(\phi_c-\sigma)^2-(\phi_c-\sigma\lambda\omega\phi_c^p)^2}{2p}, \\
\delta &= \sqrt{3}\lambda^{-1}\phi_c^{-p}\sqrt{\lambda\phi_c^p(3\lambda\phi_c^p+2p(\sigma\omega\phi_c)^{-1})}, \\
\epsilon &= -\frac{\phi_c}{3}\left(3\omega\lambda\sigma\phi_c^p + \sqrt{3\omega\lambda\sigma\phi_c^{-1+p}(2p+3\omega\lambda\sigma\phi_c^{1+p})}\right)
\ln\left[\frac{1}{\omega\lambda\phi_c^p}\right].
\end{align}

\end{document}